\documentclass{aa}

\usepackage{amsmath}
\usepackage{graphicx}
\usepackage{txfonts}
\usepackage[colorlinks]{hyperref}
\usepackage[table,dvipsnames]{xcolor}
\usepackage[most]{tcolorbox}
\usepackage[%
    cachedir=minted-cache,
    frozencache=true,
]{minted}
\tcbuselibrary{minted}
\usepackage{booktabs}
\usepackage{gensymb}

\newcommand{\bbA}{{\bf{A}}}

\newcommand{\bbI}{{\bf{I}}}

\newcommand{\bbp}{{\bf{p}}}

\newcommand{\bbr}{{\bf{r}}}

\newcommand{\bbV}{{\bf{V}}}

\newcommand{\bC}{\mathbb{C}}

\newcommand{\bR}{\mathbb{R}}
\newcommand{\bS}{\mathbb{S}}

\newcommand{\cB}{\mathcal{B}}

\newcommand{\cO}{\mathcal{O}}
\newcommand{\cP}{\mathcal{P}}

\newcommand{\cV}{\mathcal{V}}
\newcommand{\cW}{\mathcal{W}}

\newcommand{\bigBrack}[1]{\left[ #1 \right]}
\newcommand{\bigCurly}[1]{\left\{ #1 \right\}}
\newcommand{\bigParen}[1]{\left( #1 \right)}
\newcommand{\innerProduct}[2]{\left\langle #1, #2 \right\rangle}

\newcommand{\abs}[1]{\left| #1 \right|}

\DeclareMathOperator*{\argmin}{arg\,min}

\newcommand{\hvox}{HVOX}
\newcommand{\hvoxI}{\texttt{HVOX}}  
\newcommand{\pycsou}{Pycsou}  
\newcommand{\vis}{\text{vis}}
\newcommand{\pix}{\text{pix}}
\newcommand{\tNUFFTone}{NUFFT$_{1}$}
\newcommand{\tNUFFTthree}{NUFFT$_{3}$}
\newcommand{\tFINUFFT}{FINUFFT}
\newcommand{\wid}{\text{s}}
\newcommand{\mesh}{\text{mesh}}
\newcommand{\twgridder}{W-gridder}
\newcommand{\blk}{\text{blk}}
\newcommand{\hpix}{HEALPix}
\newcommand{\dcos}{DCOS}
\newcommand{\ducc}{\texttt{ducc}}
\newcommand{\nifty}{NIFTY}
\newcommand{\niftytd}{NIFTY 2D}
\newcommand{\hvoxMB}{HVOX-500MB}
\newcommand{\nmse}{NMSE}

\newtcolorbox[
  auto counter
]{algorithmBox}[2][]{
  enhanced,
  float=t!,
  colback=gray!3!white,
  colframe=gray!75!black,
  toptitle=1mm,
  fonttitle=\bfseries,
  width=\columnwidth,
  center, 
  unbreakable,
  parbox=false, 
  title={Algorithm~\thetcbcounter: #2},
  #1
}

\newtcblisting[
  auto counter
]{myListing}[3][]{%
  enhanced, 
  float=t!,
  leftupper=1mm,
  rightupper=1mm,
  colback=gray!3!white,
  colframe=green!50!black,
  toptitle=1mm,
  fonttitle=\bfseries,
  title={Listing~\thetcbcounter: #2},
  listing only,
  listing engine=minted,
  minted style=xcode,
  minted options={fontsize=\footnotesize,breaklines,autogobble,#3}, 
  #1 
}

\newcommand{\internalColor}{Mulberry}
\newcommand{\externalColor}{NavyBlue}
\hypersetup{
  citebordercolor=\internalColor,
  citecolor=\internalColor,
  filebordercolor=\externalColor,
  filecolor=\externalColor,
  linkbordercolor=\internalColor,
  linkcolor=\internalColor,
  menubordercolor=\internalColor,
  menucolor=\internalColor,
  runbordercolor=\externalColor,
  runcolor=\externalColor,
  urlbordercolor=\externalColor,
  urlcolor=\externalColor,
}

\newcommand{\secRef}[1]{\hyperref[#1]{Sect.~\ref{#1}}}
\newcommand{\figRef}[1]{\hyperref[#1]{Fig.~\ref{#1}}}
\newcommand{\tabRef}[1]{\hyperref[#1]{Table~\ref{#1}}}
\newcommand{\algoRef}[1]{\hyperref[#1]{Algorithm~\ref{#1}}}
\newcommand{\eqRef}[1]{\hyperref[#1]{Eq.~\ref{#1}}}
\newcommand{\lstRef}[1]{\hyperref[#1]{Listing~\ref{#1}}}

\begin{document}
\title{\hvox{}: Scalable Interferometric Synthesis and Analysis of Spherical Sky Maps}
\author{%
    S.~Kashani \inst{1}
    \and
    J.~Rué Queralt \inst{2}
    \and
    A.~Jarret \inst{1}
    \and
    M.~Simeoni \inst{2}
}
\institute{
    Audiovisual Communications Laboratory, École Polytechnique Fédérale de Lausanne, Switzerland \\
    \email{sepand.kashani@epfl.ch, adrian.jarret@epfl.ch}
    \and
    Center for Imaging, École Polytechnique Fédérale de Lausanne, Switzerland \\
    \email{matthieu.simeoni@epfl.ch, joan.ruequeralt@epfl.ch}
}
\date{Received 2023.06.09; accepted YYYY.MM.DD}
\abstract
{
    Analysis and synthesis are key steps of the radio-interferometric imaging process, serving as a bridge between visibility and sky domains.
    They can be expressed as partial Fourier transforms involving a large number of non-uniform frequencies and spherically-constrained spatial coordinates.
    Due to the data non-uniformity, these partial Fourier transforms are computationally expensive and represent a serious bottleneck in the image reconstruction process.
    The W-gridding algorithm achieves log-linear complexity for both steps by applying a series of 2D non-uniform FFTs (NUFFT) to the data sliced along the so-called $w$ frequency coordinate.
    A major drawback of this method however is its restriction to direction-cosine meshes, which are fundamentally ill-suited for large field of views.
}
{
    This paper introduces the \hvox{} gridder, a novel algorithm for analysis/synthesis based on a 3D-NUFFT.
    Unlike W-gridding, the latter is compatible with arbitrary spherical meshes such as the popular \hpix{} scheme for spherical data processing.
}
{
    The 3D-NUFFT allows one to optimally select the size of the inner FFTs, in particular the number of W-planes.
    This results in a better performing and auto-tuned algorithm, with controlled accuracy guarantees backed by strong results from approximation theory.
    To cope with the challenging scale of next-generation radio telescopes, we propose moreover a chunked evaluation strategy: by partitioning the visibility and sky domains, the 3D-NUFFT is decomposed into sub-problems which execute in parallel, while simultaneously cutting memory requirements.
}
{
    Our benchmarking results demonstrate the scalability of \hvox{} for both SKA and LOFAR, considering state-of-the-art challenging imaging setups. (100+ M\pix{} images, with extreme-scale baselines.)
    \hvox{} is moreover computationally competitive with W-gridder, despite the absence of domain-specific optimizations in our implementation.
    Accuracy-wise we show that, while W-gridder with interpolation from/to \hpix{} maps fails to perform error-controlled analysis or synthesis, \hvox{} can achieve relative accuracies up to $10^{-9}$ in double precision.
    Finally, we release \hvoxI{}, the first software implementation of the \hvox{} gridder.
}
{}

\maketitle

\section{Introduction}\label{sec:introduction}
Image cubes are the starting point of many scientific inquiries in astronomy, including for example source detection/classification \citep{masias2013quantitative} or deconvolution \citep{starck2000combined}.
Image production is thus a key pillar of the SKA Science Data Processor (SDP) \citep{scaife2018sdpimg}.

The typical imaging pipeline in radio-astronomy is a multi-stage process.
Two key ingredients within are notably image \emph{analysis} (A) and \emph{synthesis} (S), dual and computationally intensive operations allowing to predict visibilities from a given sky distribution, and vice versa.
These operations are commonly known as \emph{gridding} and \emph{de-gridding} in radio-astronomy, and are omnipresent in all imaging methods.
For example CLEAN and its variants perform one A\&S step every major cycle to correct for inaccuracies in the residual image accumulated over minor cycles \citep[Chap.\ 11]{thompson2017interferometry}.
Similarly, proximal algorithms and Monte Carlo methods in variational and Bayesian inference perform one A\&S step per iteration \citep{terris2023image,cai2018uncertaintyI,cai2018uncertaintyII}.
For this reason, A\&S tend to dominate the overall computational complexity of imaging tasks.
Moreover, direct evaluation of these operators has bi-linear complexity in the number of visibilities and pixels, a cost which becomes prohibitively expensive beyond small imaging problems.
For the SDP imaging pipeline to meet its ambitious imaging and scientific goals, it is paramount to have \emph{fast}, \emph{accurate} and \emph{scalable} algorithms to perform A\&S.

The radio-astronomy community has developed a rich suite of fast gridders over the years to perform A\&S.
Prominent methods include image-plane faceting \citep{kogan2009faceted}, $w$-projection \citep{cornwell2008noncoplanar}, $w$-snapshot \citep{10.1117/12.929336}, $w$-stacking \citep{offringa2014wsclean}, W-gridding\footnote{Also known as the \emph{NIFTY} gridder.} \citep{ye2022high}, Image Domain Gridding (IDG) \citep{van2018image}, and their derivatives.
By leveraging the 2D Fast Fourier Transform (FFT), the computational cost of most of these methods scales linearly with the number of visibilities and log-linearly in the number of pixels.
Currently W-gridding and IDG are considered the leading gridders to handle the W- and A-terms in the SKA respectively.

To leverage the 2D FFT, the common trait among these methods is to pixelize the sky using a \emph{direction-cosine} mesh (\dcos) \citep{thompson2017interferometry}.
However this rigid grid choice leads to \emph{computational} and \emph{scientific} issues in wide-field imaging,  described hereafter.

From a computational perspective first, current gridders perform 2D FFTs of mesh size equal to the pixel resolution, and their size is often several times larger than an interferometer's critical resolution to simplify image-processing tasks such as deconvolution at sub-Nyquist precision.
However the FFT is a global transform with high communication overhead.
As such it does not scale well to large multi-dimensional meshes.
Out-of-core FFTs exacerbate the problem when the full grid does not fit in memory due to their dependance on the I/O sub-system \citep{cormen1998performing}.
Large imaging tasks tend therefore to be FFT-bound.
Scalable image production in the SDP is thus still an open question \citep{scaife2018sdpimg}.

From a science production perspective then, \dcos{} grids are also often ill-suited for wide-field sky maps.
Indeed, the imaging pipeline rarely ends at image formation: it is followed by complex post-processing pipelines to extract secondary data products for further study.
Prominent tasks include (1) image filtering for noise reduction, feature enhancement, or image deconvolution \citep{starck1998image}; and (2) feature classification via harmonic analysis (i.e. wavelets, spherical harmonics \citep{stephane1999wavelet,rafaely2015fundamentals,starck2006wavelets}) or popular learning-based methods \citep{bronstein2017geometric,perraudin2019deepsphere}.
The former assume Euclidean signal domains with uniform sampling.
\dcos{} grids sample the sphere non-uniformly however, hence the poor performance of classical image processing tools when applied to this mesh.
Similarly, learning-based methods do not generalize well to fields where spatial statistics are non-uniform \citep{perraudin2019deepsphere,bronstein2017geometric}.
To overcome these issues, \dcos{} meshes can be re-interpolated to quasi-uniform tesselations prior to processing.
The \emph{\hpix{}} grid in particular is ideally suited to spherical image processing due to the availability of fast harmonic transforms in this mesh and good support by the FITS format \citep{gorski2005healpix,calabretta2007mapping,lanusse2012spherical}.
Exact spherical interpolation is non-trivial and expensive to perform on large images however \citep{simeoni2021functional}.
Moreover, performing approximate re-interpolation breaks any accuracy guarantees provided by gridders which produced the original image.

Such limitations stress the need for gridders which (1) work with alternative spherical meshes to ease post-processing while retaining speed and accuracy; and (2) perform FFTs where the mesh size is independent of the pixel-count to scale.

This paper introduces a new gridder dubbed \hvox{} (short for \emph{Heisenberg voxelisation}) which directly addresses the aforementioned issues.
Our main contributions are threefold:
\begin{description}
    \item[\textbf{Methodology}]
    \hvox{} performs A\&S using a \emph{type-3 non-uniform Fast Fourier Transform} (\tNUFFTthree{}) in 3D, a standard signal processing transform \citep{barnett2019parallel}.
    \tNUFFTthree{} algorithms naturally support most desired properties of gridders, namely:
    \begin{itemize}
        \item
        \emph{mesh-agnostic A\&S} via interpolating to/from a canonical 3D Heisenberg mesh of fundamental harmonics;

        \item
        \emph{guaranteed tunable accuracy} via strong results from approximation theory and window design;

        \item
        \emph{scalable} by performing 3D FFTs of mesh size \emph{independent} of the pixel count.
        The FFT dimensions are critically chosen by computing the size of "Heisenberg boxes" needed to encapsulate visibilities and the field of view.
    \end{itemize}
    \item[\textbf{Algorithmic}]
    \hvox{} introduces a hierarchical method to evaluate the type-3 transform, which drastically cuts the FFT runtime and memory requirements by partitioning visibility/sky domains into \emph{chunks}.
    Our chunk-based approach gives users direct control on the FFT memory consumed for better scalability, the ability to tailor the evaluation strategy on a per-chunk basis, and is inherently fast since chunks can be computed in parallel.
    It is particularly efficient in the context of sparse baseline configurations and/or sky maps (e.g., as encountered in CLEAN iterations or during direction-dependant calibration).
    In such contexts, our approach is up to one order of magnitude faster than \twgridder{}.
    \item[\textbf{Systems}]
    We provide \hvoxI{}, the first CPU-based implementation of the \hvox{} method.
    Built on Dask \citep{rocklin2015dask}, Pycsou \citep{matthieu_simeoni_2021_4715243} and the \tFINUFFT{} type-3 engine \citep{barnett2019parallel}, we show that \hvoxI{} is computationally competitive with \twgridder{}, despite the absence of domain-specific optimizations, and can easily scale to challenging state-of-the-art imaging setups. (E.g., 100+ M\pix{} images with all SKA-LOW baselines.)
    Our benchmarking experiments reveal moreover that, while W-gridder with interpolation from/to \hpix{} maps fails to perform error-controlled analysis or synthesis, \hvox{} can achieve relative accuracies up to $10^{-9}$ in double precision.
 \end{description}

Finally, the \hvox{} method does not require any architectural changes to the interferometric software pipeline:
it can be used as a drop-in replacement for current \dcos{} gridders, easing adoption and deployment in production environments.

This paper is structured as follows:
\secRef{sec:ana-synth} describes the analysis and synthesis steps, and surveys current evaluation strategies.
\secRef{sec:hvox} describes the \hvox{} method to perform A\&S.
Implementation details of \hvox{} are described in \secRef{sec:hvox-impl}.
\secRef{sec:results} shows simulation results comparing \hvoxI{} to \twgridder{} in terms of accuracy and computation time in various imaging setups.
Finally \secRef{sec:future} concludes with a perspective on the implications of \hvox{} for radio-interferometry.

\section{Analysis \& synthesis}\label{sec:ana-synth}
An interferometer samples the complex visibility function \citep{thompson2017interferometry}:%
\footnote{
    Antenna beamshapes have been omitted without loss of generality.
    Extending the \hvox{} method to handle direction-dependant effects (DDE) will be the subject of future work.
}
\begin{equation}
    \label{eq:measurement-equation}
    V_{i} = \int_{\bS^{2}} I(\bbr) \exp\bigParen{-j \innerProduct{\bbr}{\bbp_{i}}} \text{d}\bbr,
\end{equation}
where $I(\bbr) \in \bR_{+}$ denotes the probed spherical sky intensity map, $\bbp_{i} \in \bR^{3}$ the $i$-th wavelength-normalized baseline, and $V_{i} \in \bC$ the measured visibility.
Once discretized, \eqRef{eq:measurement-equation} gives the \emph{analysis} equation and its adjoint \emph{synthesis}:
\begin{align}
    V_{i} & = \sum_{\bbr \in \Theta_{\pix}} I(\bbr) \alpha_{\Theta}(\bbr) \exp\bigParen{-j \innerProduct{\bbr}{\bbp_{i}}} \label{eq:analysis}, \\
    \hat{I}(\bbr) & = \sum_{i = 1}^{N_{\vis}} V_{i} \alpha_{\Theta}(\bbr) \exp\bigParen{j \innerProduct{\bbr}{\bbp_{i}}} \label{eq:synthesis},
\end{align}
where $\Theta_{\pix}$ denotes some discrete tesselation of the sphere $\bS^{2}$ with size $N_{\pix}$, $\alpha_{\Theta}(\bbr) \in \bR$ some tesselation weights (for numerical quadrature), and $N_{\vis}$ the total number of  visibilities.
Note that the synthesized intensity map $\hat{I}$, generally referred to as the \emph{dirty image} (or \emph{residual image} in the context of CLEAN major cycles), differs from the probed sky intensity map and can be seen as a blurred version of the latter \citep{simeoni2019graph}.

Naive evaluation of \eqRef{eq:analysis} and \eqRef{eq:synthesis} costs $\cO\bigParen{N_{\vis} N_{\pix}}$ operations.
This bi-linear complexity in visibilities and pixel-count quickly becomes intractable.
Fast evaluation strategies seek to reduce this bi-linear complexity to (at most) log-linearity in $N_{\vis}$ and $N_{\pix}$ respectively.
Note that since analysis and synthesis form an adjoint pair, any fast algorithm computing synthesis can be reversed to obtain an equally-fast algorithm for analysis.
Without loss of generality and for the sake of clarity, we therefore limit our discussion below to synthesis.

\subsection{Synthesis evaluation strategies}\label{subsec:synth-strategies}
The core idea used by gridders today is to evaluate (parts of) \eqRef{eq:synthesis} using a 2D FFT.
Assuming sky coordinates $\Theta_{\pix}$ lie on a direction-cosine mesh on the tangent plane at the field center, \eqRef{eq:synthesis} can be rewritten as
\begin{align}
    \hat{I}[l, m]
    & \propto \sum_{i=1}^{N_{\vis}} V_{i} \exp\bigParen{j \bigBrack{u_{i} l + v_{i} m + w_{i} \bigCurly{n(l, m) - 1}}} \nonumber \\
    & = \sum_{i=1}^{N_{\vis}} V_{i} \cW_{i}(l, m) \exp\bigParen{j \bigBrack{u_{i} l + v_{i} m}}, \label{eq:synthesis-dcos-mesh}
\end{align}
where $(u_{i}, v_{i}, w_{i}) \in \bR^{3}$ are the coordinates of baseline $\bbp_{i}$ in the UVW frame, and $n(l, m) = \sqrt{1 - l^{2} - m^{2}}$.
Fast evaluation of \eqRef{eq:synthesis-dcos-mesh} depends on the significance of the so-called \emph{W-term} $\cW_{i}(l, m) = \exp\bigParen{j w_{i} \bigCurly{n(l,m) - 1}}$.

For small field-of-views, the W-term can be neglected (e.g., $\cW_{i}(l, m) \simeq 1$) such that \eqRef{eq:synthesis-dcos-mesh} identifies as computing the $(l_{j}, m_{j})_{j=1}^{N_{\pix}}$ 2D Fourier Series (FS) coefficients of the Dirac stream $\cV(u,v) = \sum_{i = 1}^{N_{\vis}} V_{i} \delta(u - u_{i}, v - v_{i})$.
The former is an instance of the \emph{type-1 non-uniform Fast Fourier Transform} problem (\tNUFFTone{}) in 2D \citep{barnett2019parallel}.
It performs a mapping from non-uniform frequency coordinates to uniform spatial coordinates.
\tNUFFTone{} transforms evaluate \eqRef{eq:synthesis-dcos-mesh} efficiently according to the following steps:
\begin{enumerate}
    \item\label{it:nufft1-spread}
    \emph{spreading} off-grid visibilities onto a uniform 2D mesh of size $N_{\pix}$ using a well-chosen 2D kernel $\phi$;

    \item
    \emph{transforming} gridded visibilities to FS coefficients via a 2D FFT;

    \item
    \emph{tapering} FS coefficients to correct the effect of step \ref{it:nufft1-spread}.
\end{enumerate}
The total arithmetic cost of an \tNUFFTone{} is hence $\cO\bigParen{N_{\vis} N_{\wid} + N_{\pix} \log N_{\pix} + N_{\pix}}$, where each term represents the cost of spreading, FFT, and tapering respectively.
The constant $N_{\wid}$ denotes the support (in number of harmonic samples) of $\phi$.
Its choice strongly affects the spreading cost.
It is common therefore to use compact-support kernels in place of the optimal Dirichlet interpolation kernel \citep{barnett2019parallel,vetterli2014foundations}.
By doing so \tNUFFTone{} becomes an approximate transform.
This is acceptable in practice since there is a class of compact-support kernels%
\footnote{
    Notable spreading kernels include \emph{prolate spheroidal wave functions} (PSWF, \cite{slepian1965some}), Gaussian \citep{dutt1993fast}, Kaiser-Bessel \citep{kuo1966system}, and the \emph{exponential of semi-circle} (ES) pulse \citep{barnett2019parallel}.
}
such that the $\epsilon$ relative error of the transform (with respect to direct evaluation of the complex exponential sum) can be controlled via the size of $N_{\wid}$.
Good kernels are such that $N_{\wid} \approx \abs{\log\epsilon}^{2}$, i.e.\ the kernel size in each dimension roughly corresponds to the number of desired significant digits.

\begin{figure}[t]
    \centering
    \includegraphics[width=\linewidth]{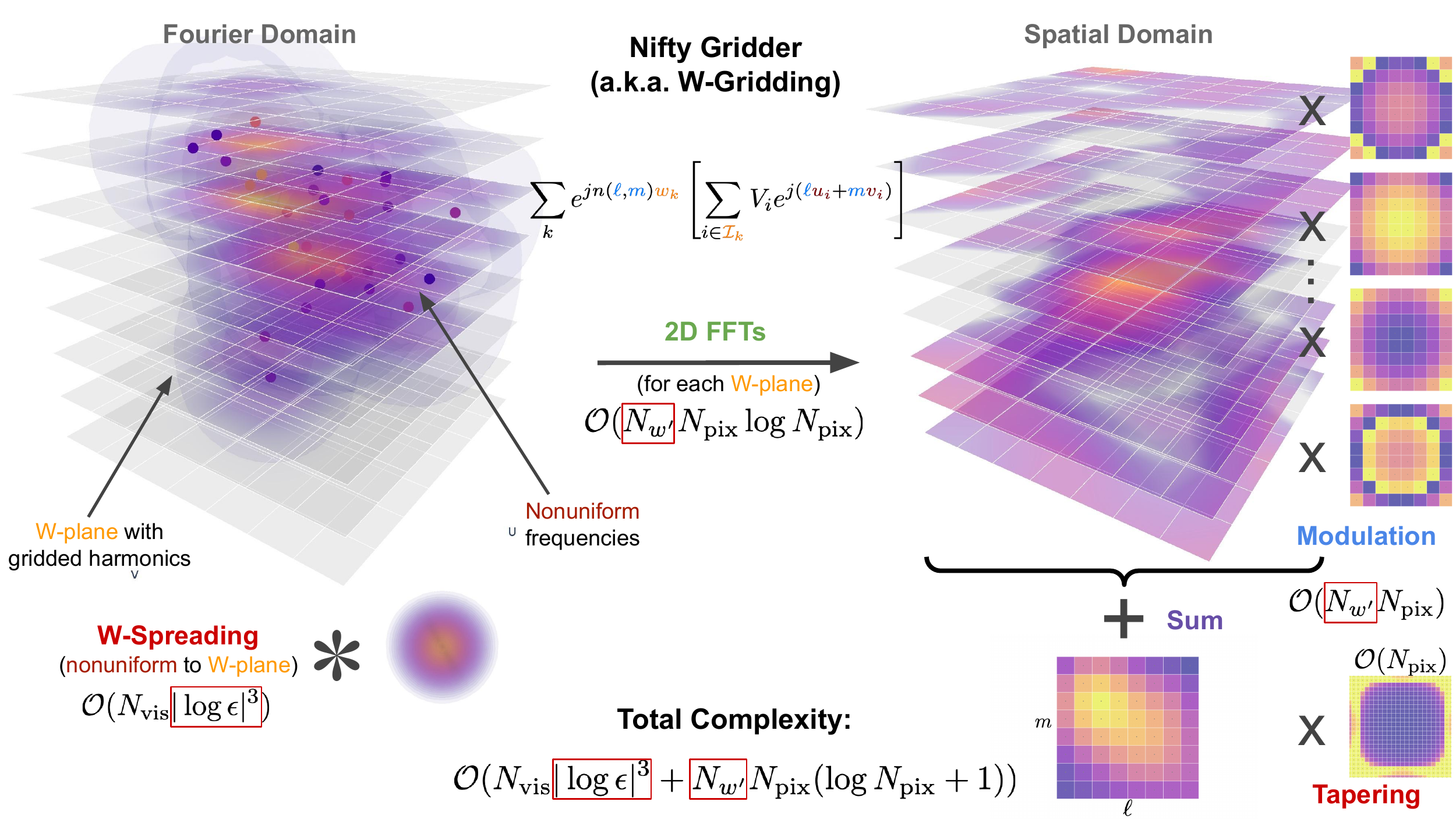}
    \caption{%
        Schematic view of W-gridding for synthesis.
        Visibilities are gridded across W-planes via convolution with a 3D spreading kernel.
        Each W-plane then undergoes a 2D FFT.
        Finally, W-planes are modulated to account for their respective W-terms prior to being collapsed.
    }
    \label{fig:nifty-gridder}
\end{figure}

For large field-of-views however, the W-term cannot be neglected anymore, and evaluating \eqRef{eq:synthesis-dcos-mesh} becomes more complex.
The line of research instigated by $w$-stacking \citep{offringa2014wsclean} is to compute \eqRef{eq:synthesis-dcos-mesh} hierarchically.
In essence, visibilities are (1) binned into $N_{w}$ stacks where the W-term is roughly baseline-independent.
Each visibility stack is then (2) processed by \tNUFFTone{} as described above.
Finally (3) the stacks are linearly combined after accounting for their respective W-terms via modulation.
\twgridder{} \citep{ye2022high} extends the $w$-stacking method by spreading visibilities over multiple W-planes.
By doing so it achieves a lower error rate than $w$-stacking while using less W-planes $N_{w'} < N_{w}$.
Moreover $N_{w'}$ is known analytically.
The full W-gridding process is illustrated in \figRef{fig:nifty-gridder}.
\begin{figure}[t]
    \centering
    \includegraphics[width=\linewidth]{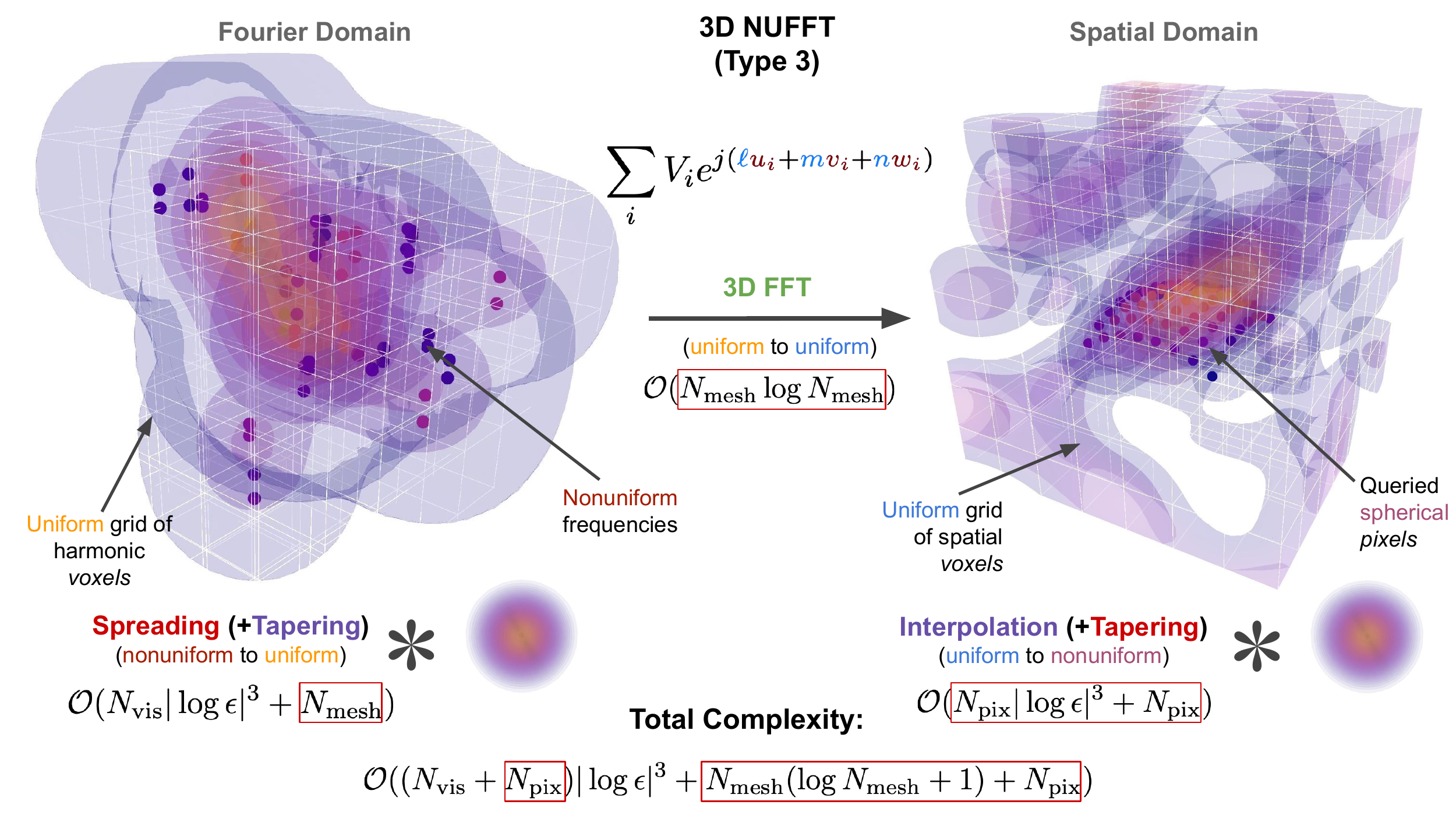}
    \caption{%
        Schematic view of the type-3 NUFFT in 3D for synthesis.
        Visibilities are spread onto a uniform grid of harmonic voxels.
        These are transformed to spatial harmonics via a 3D FFT, followed by interpolation onto off-grid sky coordinates.
        Finally, spherical pixels are tapered to correct the effect induced by spreading.
    }
    \label{fig:nufft3}
\end{figure}

The arithmetic complexity of \twgridder{} is $\cO\bigParen{N_{\vis} \abs{\log\epsilon}^{3} + N_{w'} N_{\pix} \log N_{\pix} + N_{w'} N_{\pix}}$.
Compared to \tNUFFTone{}, its spreading cost is higher due to performing 3D gridding instead of 2D.
Moreover FFT and modulation costs scale proportionally to the number of layers $N_{w'}$.

In both cases, we observe that \dcos{} meshes fundamentally shape the way current gridders perform analysis and synthesis.
The regular structure of the mesh indeed enables 2D FFTs to break away from the $\cO\bigParen{N_{\vis} N_{\pix}}$ complexity of direct evaluation.
However, as outlined in \secRef{sec:introduction}, performing $N_{\pix}$-sized FFTs (with $N_{\pix}$ often much larger than Nyquist's critical resolution) scales poorly to requirements of the SDP.
Moreover, as already discussed, \dcos{} meshes are problematic for processing wide-field spherical sky maps.

\section{The \hvox{} method}\label{sec:hvox}
Current gridders all treat the $w$-coordinate separately from the $(u,v)$-coordinates (e.g., via layered evaluation strategies such as $w$-stacking) which seems somewhat artificial.
W-gridding's idea to spread in 3 dimensions proved very insightful: by treating all dimensions symmetrically, lower error rates were achieved.
Nevertheless W-gridding retains a firm view of synthesis as a 2D problem by evaluating FFTs in planar fashion.
Computational capacity notwithstanding, there may be merit in dropping special treatment of the W-term entirely.
This can be achieved by considering synthesis from the ground-up starting from \eqRef{eq:synthesis} rather than the \dcos-based variant (\eqRef{eq:synthesis-dcos-mesh}).

The key insight is to identify \eqRef{eq:synthesis} as an instance of the \emph{type-3 non-uniform Fast Fourier Transform} (\tNUFFTthree{}) in 3D \citep{barnett2019parallel}, i.e.\ the problem of mapping non-uniform baseline coordinates $\bbp_i\in\bR^{3}$ to (potentially) non-uniform sky coordinates $\bbr\in\Theta_{\pix}\subset\bS^{2}$.
\tNUFFTthree{} performs synthesis by interleaving type-1 and type-2\footnote{A type-2 NUFFT is the adjoint operation to a type-1: it maps uniform harmonics to non-uniform coordinates.} transforms.
Concretely, \tNUFFTthree{} algorithms evaluate \eqRef{eq:synthesis} by:
\begin{enumerate}
    \item
    \emph{spreading} non-uniform visibilities onto a uniform 3D Heisenberg mesh of size $N_{\mesh}$.
    Spreading is done using a well-chosen 3D kernel $\phi$, followed by tapering;

    \item
    \emph{transforming} gridded visibilities to spatial harmonics via a 3D FFT;

    \item
    \emph{interpolating} harmonics to non-uniform sky coordinates via a well-chosen 3D kernel $\hat{\phi}$;

    \item
    \emph{tapering} pixels to correct the effect of spreading/windowing.
\end{enumerate}
The full \tNUFFTthree{} process is illustrated in \figRef{fig:nufft3}.
The arithmetic complexity of \tNUFFTthree{} methods is given by
\begin{equation*}
    \cO\bigParen{\bigBrack{(N_{\vis} + N_{\pix}) \abs{\log\epsilon}^{3} + (N_{\mesh} + N_{\pix})} + N_{\mesh} \log N_{\mesh}},
\end{equation*}
where $N_{\vis}, N_{\pix}, \epsilon$ are as for \tNUFFTone{}, and $N_{\mesh}$ is defined via \citep{barnett2019parallel}
\begin{align}
    N_{\mesh} & \propto \cV = \prod_{k=1}^{3} \cB_{k} \cP_{k} \label{eq:type3-mesh-size} \\
    \cB_{k} & = \max_{i=1,\ldots,N_{\vis}} \bigBrack{\bbp_{i}}_{k} - \min_{i=1,\ldots,N_{\vis}} \bigBrack{\bbp_{i}}_{k} \label{eq:cb-def} \\
    \cP_{k} & = \max_{i=1,\ldots,N_{\pix}} \bigBrack{\bbr_{i}}_{k} - \min_{i=1,\ldots,N_{\pix}} \bigBrack{\bbr_{i}}_{k}. \label{eq:cp-def}
\end{align}
$\cB_k $ and $\cP_k$ can be interpreted as the size in each dimension of the bounding boxes of all baselines and pixels respectively.

The similarity between \twgridder{} and \tNUFFTthree{} is striking.
The main differences are (1) performing one 3D FFT instead of $N_{w'}$ 2D FFTs; and (2) interpolation post-FFT to target arbitrary pixel tesselations.
Both methods are thus intimately linked: despite different derivations from \eqRef{eq:synthesis} and \eqRef{eq:synthesis-dcos-mesh}, \twgridder{} is in effect trying to approximate a 3D \tNUFFTthree{} via stacked 2D FFTs.

Using a type-3 NUFFT for synthesis has however several advantages over type-1 based methods used in gridders today:
\begin{itemize}
    \item
    Interpolation from the spatial Heisenberg mesh allows one to synthesize sky maps onto arbitrary grids.
    Images can be produced directly in a tesselation suitable for image-processing pipelines without resorting to ad hoc and error-inducing re-interpolation steps.
    In essence \emph{mesh-agnostic} synthesis solves all qualitative issues raised in \secRef{sec:introduction}.
    \item
    Similar to \tNUFFTone{}, \tNUFFTthree{} approximates \eqRef{eq:synthesis} due to the use of compact-support kernels in the spreading/interpolation steps.
    However the aliasing error induced by not using the Dirichlet kernel is well-understood.
    In particular strong aliasing results exist in the signal processing literature \citep{barnett2021aliasing}.
    Compared with engineered gridders where analytical error bounds may be unavailable, \tNUFFTthree{} can auto-choose a suitable kernel support required to achieve a desired accuracy for "typical" signals \citep[Sec.\ 4]{barnett2019parallel}.
    \item
    The 3D FFT is performed at mesh size $N_{\mesh}$, which is independent of $N_{\pix}$.
    It is a function of $\cV$, the visibility/sky-domain volume product. (See \eqRef{eq:type3-mesh-size}.)
    Moreover the box sizes are chosen as small as possible based on the \emph{Heisenberg uncertainty principle} (HUP).
    The 3D FFT is hence performed at the critical Nyquist resolution.
    Due to the link with the HUP, visibility/sky bounding boxes are colloquially called \emph{Heisenberg boxes}. \\
\end{itemize}
With an arithmetic intensity linear in $(N_{\vis}, N_{\pix})$, and an FFT cost independent of image resolution, evaluating synthesis via \tNUFFTthree{} should scale well.
From a methodology perspective therefore, using \tNUFFTthree{} algorithms to perform analysis and synthesis seems ideal.

\subsection{Scaling to large baselines \& field-of-views}\label{subsec:hvox-scale}
In practice straightforward application of \tNUFFTthree{} to wide-field synthesis/analysis does not scale well for two reasons.

\begin{figure}[t]
    \resizebox{\hsize}{!}{\includegraphics{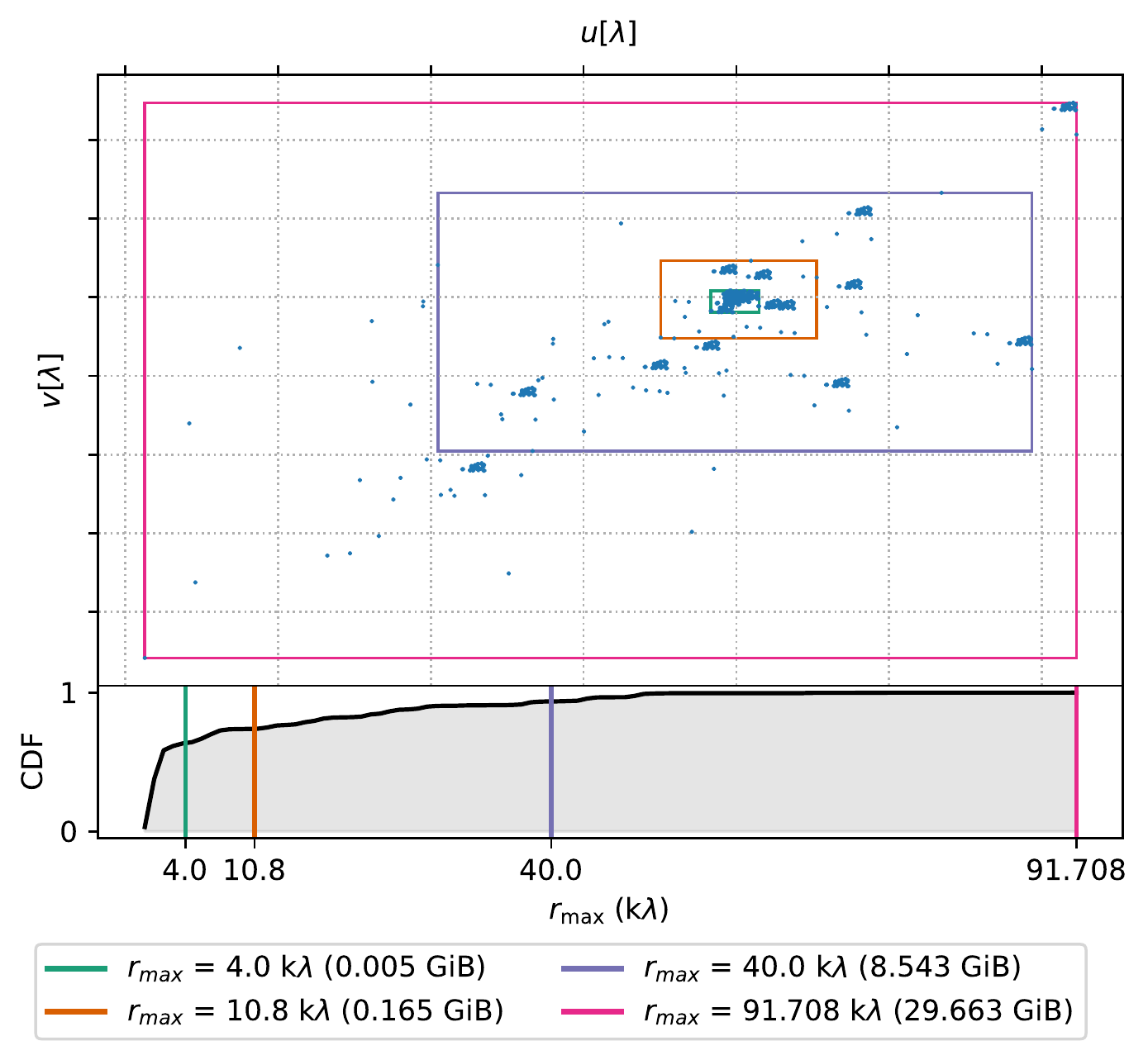}}
    \caption{
        \textbf{Top:} UV coverage of LOFAR-HBA (24 core stations \& 14 remote stations) during one short-time interval. \\
        \textbf{Bottom:} distribution of baseline counts within a maximum $L_{\infty}$ radius. \\
        Most baselines lie in a dense cluster around the origin, yet the FFT memory consumption is determined solely by the farthest baselines. (See legend.)
        The large FFT mesh is hence mostly void of spreaded visibilities.
    }
    \label{fig:hbox-dist}
\end{figure}

Firstly, the mesh size in each dimension is related to the max baseline separation and field width.
Crucially, it is independent of the number of measured visibilities and final image size.
The memory required to store the mesh, and the FFT cost to process it, may therefore grow arbitrarily large irrespective of the size of the inputs. (See \figRef{fig:hbox-dist}.)

Secondly, assuming the 3D Heisenberg mesh fits in memory, the amount of useful work performed by the 3D FFT is low due to the geometrical structure of baselines and sky tesselations.
Indeed, baselines are distributed unevenly in the UVW frame: there is a high density of short baselines near the origin, and sparse coverage in peripheral regions. (See \figRef{fig:hbox-dist}[bottom].)
Spreading therefore densely populates the "core" of the Heisenberg mesh, whereas its edges are largely void of data.
Most of the FFT effort is thus wasted given low fill-in of the mesh.
Similarly interpolating to arbitrary spherical coordinates from a large 3D Heisenberg voxel mesh is inefficient since only spatial harmonics in a tubular region surrounding the target pixels in the mesh actually contribute towards their value.
In essence, \tNUFFTthree{} pays the high arithmetic and memory cost of a large 3D FFT to process a sparse-filled mesh, spatial harmonics of which are also poorly used.

\begin{table*}[t!]
    \caption{%
      Arithmetic intensity of different gridders encountered in this paper.
      The interpolation cost of \tNUFFTone{} and \twgridder{} corresponds to tapering and modulation by the W-term respectively.
      Spreading and interpolation intensity scales linearly with the desired accuracy $\epsilon$ of the transforms.
      Due to implementation choices (see \secRef{sec:hvox-impl}), \hvoxI{} (v1) currently incurs a spread/interpolation overhead which grows with the number of chunks.
    }
    \label{tb:complexity}
    \centering
    \begin{tabular}{lccc}
        \hline\hline
        Method & Spread & FFT & Interpolation \\
        \hline
        \tNUFFTone{}         & $N_{\vis} \abs{\log\epsilon}^{2}$                                              & $N_{\pix} \log N_{\pix}$                                             & $N_{\pix}$ \\
        W-gridding           & $N_{\vis} \abs{\log\epsilon}^{3}$                                              & $N_{w'} N_{\pix} \log N_{\pix}$                                      & $N_{w'} N_{\pix}$ \\
        \tNUFFTthree{}       & $N_{\vis} \abs{\log\epsilon}^{3} + N_{\mesh}$                                  & $N_{\mesh} \log N_{\mesh}$                                           & $N_{\pix} \abs{\log\epsilon}^{3} + N_{\pix}$ \\
        \hvox{} (theory)     & $N_{\vis} \abs{\log\epsilon}^{3} + N_{\mesh}^{'} N_{\vis}^{\blk}$              & $N_{\vis}^{\blk} N_{\pix}^{\blk} N_{\mesh}^{'} \log N_{\mesh}^{'}$   & $N_{\pix} \abs{\log\epsilon}^{3} + N_{\pix}$ \\
        \hvoxI{} (v1) & $N_{\pix}^{\blk} N_{\vis} \abs{\log\epsilon}^{3} + \sum_{i,j} N_{\mesh}^{i,j}$ & $\sum_{i,j} N_{\mesh}^{i,j} \log N_{\mesh}^{i,j}$                    & $N_{\vis}^{\blk} N_{\pix} \abs{\log\epsilon}^{3} + N_{\pix} N_{\vis}^{\blk}$ \\
        \hline
    \end{tabular}
\end{table*}

The memory footprint and compute overhead of 3D synthesis methods is well known since \citep{osti_6692806}, reason why their use was quickly abandoned in favor of 2D approaches in use today.
\hvox{} overcomes the large 3D FFT problem by leveraging the baseline/tesselation structure to evaluate \eqRef{eq:synthesis} via an efficient chunking approach.
This process is best explained by viewing synthesis as matrix-vector product $\bbI = \bbA_{\Theta, \Omega} \bbV$, where
\begin{align*}
    \bbV & = \bigBrack{V_{1}, \ldots, V_{N_{\vis}}} \in \bC^{N_{\vis}}, \\
    \bbI & = \bigBrack{I(\bbr_{1}), \ldots, I(\bbr_{N_{\pix}})} \in \bR_{+}^{N_{\pix}}, \\
    \Omega & = \bigBrack{\bbp_{1}, \ldots, \bbp_{N_{\vis}}}, \\
    \Theta & = \bigBrack{\bbr_{1}, \ldots, \bbr_{N_{\pix}}},
\end{align*}
and $\bbA_{\Theta, \Omega} \in \bC^{\abs{\Theta} \times \abs{\Omega}}$ is the type-3 matrix mapping baselines $\Omega$ to sky pixels $\Theta$.

\begin{figure}[t]
    \resizebox{\hsize}{!}{\includegraphics{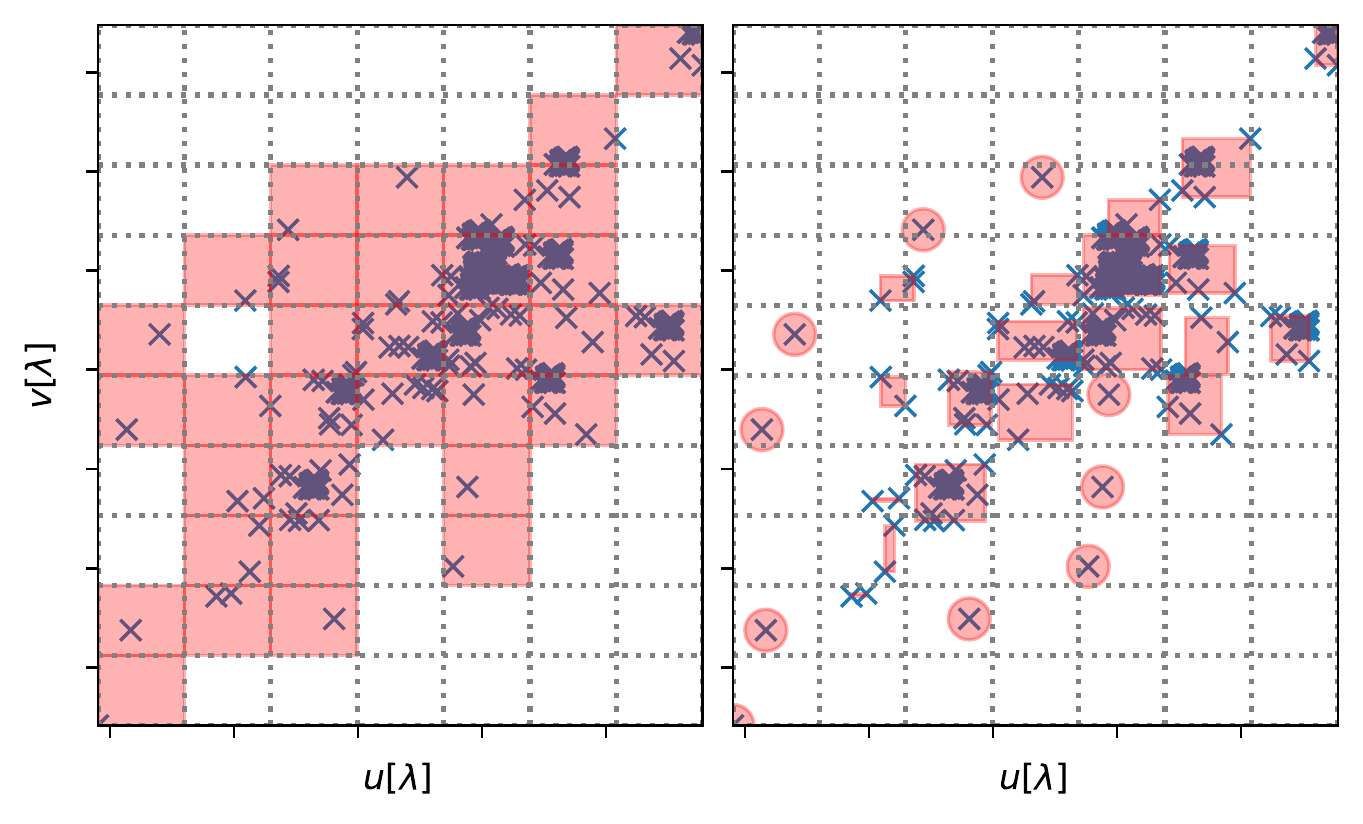}}
    \begin{center}
        \resizebox{0.55\hsize}{!}{\includegraphics{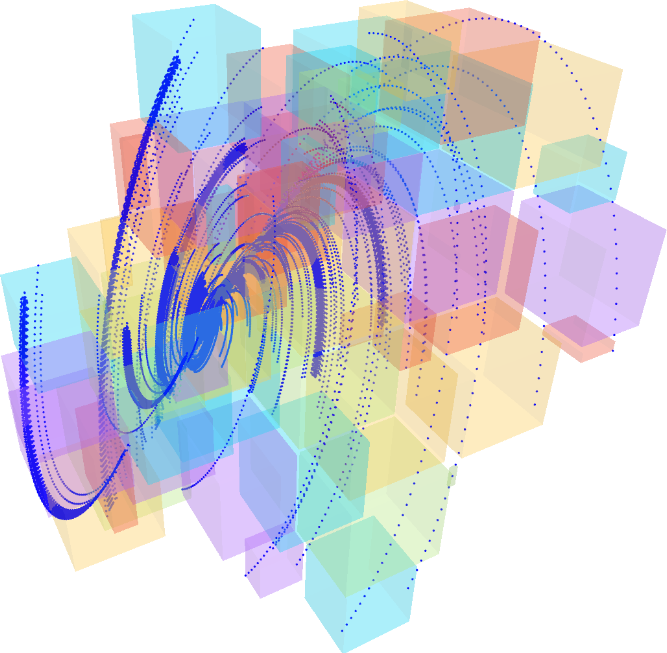}}
    \end{center}
    \caption{
        UV coverage of LOFAR-HBA (24 core stations \& 14 remote stations) during one short-time interval.
        Gridlines denote the boundaries of lattice-aligned Heisenberg volumes where associated FFT meshes consume at most $B$ bytes each. \\
        \textbf{Left:} by partitioning visibility space $\Omega$ into chunks, \hvox{} allocates FFT memory solely in regions of $\Omega$ where useful spread/FFT compute take place.
        Compared to a monolithic FFT, memory requirements of chunked evaluation is greatly reduced, and the compute-to-memory ratio significantly increased.
        Sub-transforms can moreover be computed in parallel.
        The sparse Heisenberg lattice (red boxes) is obtained using \algoRef{algo:partition}[steps \ref{it:alg-partition-1}--\ref{it:alg-partition-3}]. \\
        \textbf{Right:} effect of volume fusion, i.e.\ step \ref{it:alg-partition-4} of \algoRef{algo:partition}.
        Recursive fusion drastically cuts the number of chunks needed to cover $\Omega$ by moving volumes \emph{off} the uniform Heisenberg lattice.
        Sub-volumes with a unique visibility are depicted with red circles instead of rectangles: sub-transforms involving these Heisenberg boxes are good candidates for direct evaluation. \\
        \textbf{Bottom:} UVW coverage of LOFAR-HBA (24 core stations \& 14 remote stations) during an 8 hour observation.
        Colored boxes represent the Heisenberg boxes associated with the 61 sub-transforms obtained after auto-chunking the visibility domain $\Omega$ using \algoRef{algo:partition}.
        The box dimensions were computed using \algoRef{algo:box-dim} with $B=$ 16 MB and $\alpha = 1$.
        $\Omega$ is clearly chunked in all coordinates.
    }
    \label{fig:hbox-split}
\end{figure}

Let $\bigCurly{\Theta_{i}}_{i=1}^{N_{\pix}^{\blk}}$ and $\bigCurly{\Omega_{j}}_{j=1}^{N_{\vis}^{\blk}}$ denote some disjoint partitioning of $(\Theta, \Omega)$, and $\bbV_{\Omega_{j}}$ denote the subset of visibilities associated to baselines $\Omega_{j}$.
Then $\bbI$ can be rewritten as
\begin{equation}
    \label{eq:synthesis-matvec-split}
    \bbI_{\Theta_{i}}
    = \sum_{j=1}^{N_{\vis}^{\blk}} \bbA_{\Theta_{i},\Omega_{j}} \bbV_{\Omega_{j}},
    \quad i = 1,\ldots,N_{\pix}^{\blk},
\end{equation}
where $\bbI_{\Theta_{i}}$ corresponds to the restriction of $\bbI$ over the set $\Theta_i$.
\eqRef{eq:synthesis-matvec-split} shows that a monolithic type-3 transform mapping $\Omega$ to $\Theta$ can be computed by linear combination of $N_{\vis}^{\blk} N_{\pix}^{\blk}$ type-3 sub-transforms mapping $\Omega_{j}$ to $\Theta_{i}$.
Evaluating synthesis via \eqRef{eq:synthesis-matvec-split} in place of $\bbA_{\Theta,\Omega} \bbV$, i.e.\ a monolithic \tNUFFTthree{}, provides the following benefits:
\begin{itemize}
    \item
    Sub-transforms
    $\bigCurly{\bbA_{\Theta_{i}, \Omega_{j}}}_{i=1}^{N_{\pix}^{\blk}}$ perform the same spreading operation from off-grid visibilities to the Heisenberg mesh.
    Similarly sub-transforms
    $\bigCurly{\bbA_{\Theta_{i}, \Omega_{j}}}_{j=1}^{N_{\vis}^{\blk}}$
    perform the same interpolation operation from spatial harmonics to off-grid pixels.
    These overlapping steps can be factored out of sub-transforms to avoid redundant computations.
    As such there is \emph{no spread/interpolation overhead associated with chunked evaluation}.

    \item
    Each sub-transform operates on a subset of visibilities and pixels.
    If $(\Theta, \Omega)$ are partitioned into compact subsets, then the Heisenberg boxes associated to each sub-transform shrink drastically, and so do their respective FFT meshes.
    Due to their small size, the latter become densely populated with spreaded visibilities.
    Furthermore most spatial harmonics are queried to interpolate off-grid pixels.
    Partitioned evaluation effectively replaces a large-but-sparse $N_{\mesh}$-sized 3D FFT with many small FFTs of size $N_{\mesh}' \ll N_{\mesh}$, where sub-meshes only cover the non-zero regions of the original Heisenberg mesh.
    (See \figRef{fig:hbox-split}[left].)
    Chunked evaluation thus leads to \emph{drastic reduction of FFT memory/compute requirements} by not wasting arithmetic on empty baseline/sky regions.

    \item
    Since the FFT memory required per sub-transform scales with $\cV$, i.e.\ the volume of its Heisenberg box, the chunking process can be guided so as to form sub-transforms with capped volumes.
    Chunked evaluation therefore provides \emph{direct control of the maximum FFT memory budget per sub-transform}.

    \item
    If sub-transforms are small and independent, they can be computed in parallel to reduce synthesis time.
    Moreover the $N_{\vis}^{\blk}$ sub-transforms contributing to the same portion of the sky need only synchronize post-FFT to add their outputs to the final image grid.
    Provided $N_{\vis}^{\blk}$ is small, i.e. $(\Theta, \Omega)$ are well-partitioned, then assembly time will be low.
    Concretely, chunking (when done judiciously) allows \emph{parallel execution of sub-transforms}, with low assembly overhead.

    \item
    If $\abs{\Theta_{i}} \times \abs{\Omega_{j}}$ is small, i.e.\ sub-transform $(i,j)$ is transforming a region in baseline-space and/or sky-space with low density, then $\bbA_{\Theta_{i}, \Omega_{j}}$ is a small matrix.
    Below a certain size, it is faster to evaluate $\bbA_{\Theta_{i},\Omega_{j}} V_{\Omega_{j}}$ via direct summation rather than a type-3 NUFFT.
    Chunked evaluation therefore allows one to \emph{tailor the synthesis process at the block level} based on its characteristics.
\end{itemize}
Taken together, the properties above enable \hvox{} to perform \tNUFFTthree{}-based scalable analysis and synthesis on any domain.
The complexity of \hvox{} and other gridders surveyed is summarized in \tabRef{tb:complexity}.

\subsection{Finding a good partition}\label{subsec:find-good-partition}
\begin{algorithmBox}[label=algo:box-dim]{Find maximum Heisenberg dimensions}
    Given a maximum FFT memory budget $B > 0$ [bytes], the maximum allowed Heisenberg dimensions $(h_{k}, \eta_{k})$ in visibility/sky domains are solutions of:
    \begin{equation}
        \argmin_{h_{k}, \ \eta_{k} \; > \; 0}
        \
        N_{c}
        =
        \underbrace{\prod_{k=1}^{3} \frac{\cB_{k}}{h_{k}}}_{\text{visibility partition count}}
        \times
        \underbrace{\prod_{k=1}^{3} \frac{\cP_{k}}{\eta_{k}}}_{\text{sky partition count}}
    \end{equation}
    subject to constraints:

    \begin{enumerate}
        \item
        FFT sub-problem memory constraint \citep{barnett2019parallel}:
        \begin{equation}
            \prod_{k=1}^{3} \eta_{k} h_{k}
            \le
            \frac{8 \pi^{3} B}{\upsilon^{3} \delta},
        \end{equation}
        where $\upsilon > 1$ denotes the NUFFT's grid upsampling factor, and $\delta > 0$ the byte-size of a complex number.

        \item
        Partitions are non-degenerate/trivial respectively:
        \begin{align}
            0 & \le h_{k} \le \cB_{k}, \quad k=\{1,\ldots,3\} \\
            0 & \le \eta_{k} \le \cP_{k}, \quad k=\{1,\ldots,3\}
        \end{align}

        \item\label{it:anisotropy-condition}
        Limit cell anisotropy $\alpha \ge 1$ in visibility/sky domains:\footnote{%
            \tiny
            Condition \ref{it:anisotropy-condition} ensures Heisenberg dimensions in visibility and sky domains have similar ratios.
            Without this condition, box dimensions vary significantly with the LP solver used.
        }
        \begin{align}
            \frac{1}{\alpha} & \le \frac{h_{k}}{h_{j}} \frac{\cB_{j}}{\cB_{k}} \le \alpha, \quad k \ne j, \\
            \frac{1}{\alpha} & \le \frac{\eta_{k}}{\eta_{j}} \frac{\cP_{j}}{\cP_{k}} \le \alpha, \quad k \ne j \\
            \frac{1}{\alpha} & \le \frac{h_{k}}{\eta_{j}} \frac{\cP_{j}}{\cB_{k}} \le \alpha, \quad (j, k) = \{1,\ldots,3\}^{2}.
        \end{align}
    \end{enumerate}
\end{algorithmBox}
\hvox{}'s ability to scale hinges on finding a good partition of $(\Theta,\Omega)$ such that sub-problems do useful computations overall, i.e.\ that Heisenberg meshes be small and densely-filled.
This in turn implies Heisenberg box dimensions $(\cB_{j}, \cP_{i})$ for each sub-problem $(i,j)$ must be chosen with care.
Indeed, the distribution of visibilities/pixels, hence their Heisenberg volume, determines the compute effort and memory requirements of each sub-block.
If too small, then the compute complexity of \hvox{} tends towards $\cO\bigParen{N_{\vis} N_{\pix}}$ since sub-blocks are evaluated via direct summation below a certain threshold.
If too large, then FFT meshes have poor fill-in and computations are wasted.
Choosing dimensions "somewhere in the middle" is also sub-optimal since the overhead of assembling sub-blocks is higher than necessary.
Optimal Heisenberg dimensions seek therefore to balance:
\begin{enumerate}
    \item
    the FFT mesh size within memory constraints, while

    \item
    keeping the number of sub-blocks low to minimize the cost of block assembly, and

    \item\label{it:balanced-blocks}
    ensuring the number of visibilities in each sub-block is balanced to have similar execution times, avoiding stalls before block assembly.
\end{enumerate}
\algoRef{algo:box-dim} shows how to cast these objectives as a $6$-variable convex optimization problem which is easily solved using a linear program (LP) \citep{boyd2004convex}.
Computing good Heisenberg dimensions $(h, \eta)$ is thus cheap and fast.
Note however that \algoRef{algo:box-dim} is a data-independent method, hence cannot guarantee condition \ref{it:balanced-blocks} above holds: sub-problems $(\Theta_{i},\Omega_{j})$ thus obtained may have unbalanced runtimes.
In practice the \tNUFFTthree{} library used in \hvoxI{} solves this issue automatically.
(See \secRef{sec:hvox-impl}.)

Given $h_{k} \ll \cB_{k}$ and $\eta_{k} \ll \cP_{k}$, \hvox{} then proceeds to partition $(\Theta,\Omega)$ into \emph{chunks} $(\Theta_{i},\Omega_{j})$ via two-stage agglomerative clustering described in \algoRef{algo:partition}.%
\footnote{The same procedure is applicable to chunk sky pixels $\Theta$.}
Steps \ref{it:alg-partition-1}--\ref{it:alg-partition-3} are depicted in \figRef{fig:hbox-split}[left].
The effect of step \ref{it:alg-partition-4} is illustrated in \figRef{fig:hbox-split}[right \& bottom].

\begin{algorithmBox}[label=algo:partition]{Chunk visibilities $\Omega$}
    \begin{align*}
        \text{\textbf{in}:} \quad & \Omega = \bigParen{\bbp_{1},\ldots,\bbp_{N_{\vis}}} \in \bR^{3} \quad \text{(full dataset)} \\
                            & h = \bigBrack{h_{1}, h_{2}, h_{3}} \in \bR_{+}^{3} \quad  \text{(max Heisenberg dimensions)} \\
        \text{\textbf{out}:} \quad & \bigCurly{\Omega_{j}}_{j=1}^{N_{\vis}^{\blk}}  \qquad\text{(partition of the dataset)} \\
                             & \Omega_{1} \cup \cdots \cup \Omega_{N_{\vis}^{\blk}} = \Omega
    \end{align*}
    \begin{enumerate}
        \item\label{it:alg-partition-1}
        partition Heisenberg box of $\Omega$ into a uniform-lattice of $h$-sized volumes.
        \item\label{it:alg-partition-2}
        allocate $\bigCurly{\bbp_{1}, \ldots, \bbp_{N_{\vis}}}$ to the volumes in which they reside.
        \item\label{it:alg-partition-3}
        discard empty volumes.
        \item\label{it:alg-partition-4}
        Recursively fuse volumes $i$ and $j$ if the Heisenberg box surrounding them has dimensions less than $h$.
    \end{enumerate}
\end{algorithmBox}

\section{The \hvoxI{} implementation}\label{sec:hvox-impl}
Our implementation of \hvox{} is split into 2 packages:
\begin{itemize}
    \item a low-level and general-purpose implementation of a chunked-\tNUFFTthree{} as part of \pycsou{}, an open-source computational imaging framework for Python \citep{matthieu_simeoni_2021_4715243},
    \item a high-level \hvoxI{} Python package providing \texttt{vis2dirty()} and \texttt{dirty2vis()} functions for single-use transforms specific to radio-astronomy.
    The \hvoxI{} API mimics the same-named functions from the \ducc{} library which hosts the official implementation of \twgridder{} \citep{arras2021efficient}.
    \texttt{vis2dirty()} and \texttt{dirty2vis()} are just thin wrappers around our chunked-\tNUFFTthree{} implementation in \pycsou{}.
\end{itemize}
Without loss of generality, we detail below the low-level chunked-\tNUFFTthree{} implementation, which is suitable for batched evaluation of analysis and synthesis steps.

The choice of splitting the implementation is motivated by the fact that chunked evaluation described in \secRef{sec:hvox}, which is the backbone of the \hvox{} method, is a general approach to compute \tNUFFTthree{}-s by leveraging structure in the frequency/spatial domains.
The chunked-\tNUFFTthree{} is thus applicable to more contexts than just radio-astronomy.
It is hence distributed seperately as part of \pycsou{} to allow its re-use across imaging modalities, notably in MRI and tomography where NUFFT algorithms are cornerstone.
Moreover our implementation benefits from some built-in features of this library, in particular:
\begin{itemize}
    \item Support for distributed and out-of-core computing on CPUs/GPUs via Dask \citep{rocklin2015dask}.
    \item A precision context manager for changing locally the compute precision.
    \item Vectorized operators to efficiently process multiple inputs in parallel.
\end{itemize}

Note that we have implemented the chunked-\tNUFFTthree{} in \pycsou{} without any domain-specific optimizations so as to maximize portability.
To do so, we limit dependencies to those of \pycsou{}, namely NumPy, SciPy, Dask,%
\footnote{
    The choice of Dask is in part motivated by efforts such as Dask-MS which enable task-oriented imaging pipelines in radio-astronomy \citep{perkins2021dask}.
    More broadly, Dask is a candidate execution framework for the SDP \citep{allan2019dask}.
} and Numba.\footnote{Numba is only used in the \texttt{eps=0} case, which corresponds to performing analysis and synthesis via direct evaluation. (See \lstRef{lst:pycsou_example}.)}
Moreover we use the \texttt{finufft} package to perform all NUFFT transforms involved in sub-problems \citep{barnett2019parallel}.
This C++ library was chosen as it provides high-speed, high-precision, low-memory domain-agnostic NUFFT transforms.
Despite being (mostly) Python-backed, our implementation is competitive for both small and large problems.
(See benchmarking results in \secRef{sec:results}.)

\lstRef{lst:pycsou_example} shows a typical chunked-NUFFT instantiation to perform analysis and synthesis, composed of 3 steps:
\begin{enumerate}
    \item\label{it:pycsou-ex-step-instantiation} \texttt{NUFFT.type3()}: instantiation of a \tNUFFTthree{} transform,
    \item\label{it:pycsou-ex-step-chunking} \texttt{[auto\_chunk|allocate]()}: chunking of visibility/sky distributions,
    \item\label{it:pycsou-ex-step-transform} \texttt{[apply|adjoint]()}: application of chunked analysis \& synthesis transforms.
\end{enumerate}

\begin{myListing}[label=lst:pycsou_example,minted language=python]
    {Analysis \& Synthesis via \pycsou}
    {}
    from pycsou.runtime import Precision, Width
    from pycsou.operator.linop import NUFFT

    with Precision(Width.SINGLE):
      A = NUFFT.type3(   # analysis operator
            x=pix,            # (N_pix, 3)
            z=uvw / _lambda,  # (N_vis, 3)
            isign=-1,
            real=True,
            chunked=True,
            parallel=True,
            eps=<float>,      # transform accuracy
            ntrans=<int>,
            nthreads=<int>,
      )
      # auto-determine a good x/z chunking strategy
      x_chunks, z_chunks = A.auto_chunk()

      # apply the chunking strategy.
      A.allocate(x_chunks, z_chunks)

      V = A.apply(I)  # i.e., dirty2vis()
      I = A.adjoint(V)  # i.e., vis2dirty()
\end{myListing}

The \texttt{type3()} constructor allows the instantiation of monolithic \tNUFFTthree{} transforms and our chunked implementation:
the former directly creates a \texttt{finufft.Plan} to perform the \tNUFFTthree{} transform, whereas setting \texttt{chunked=True} delays planning until runtime. (See below.)
Moreover a \tNUFFTthree{} operator can be created in either analysis mode as seen in \lstRef{lst:pycsou_example}, or in synthesis mode with \texttt{x/z} reversed.
The advantage of analysis mode however is that the adjoint can be enforced to be real-valued via the \texttt{real} parameter, which is the case for sky intensity maps.

The \texttt{auto\_chunk()} method, i.e.\ stage \ref{it:pycsou-ex-step-chunking}, consists in running \algoRef{algo:box-dim} and \algoRef{algo:partition} via calls to SciPy routines.
The $(\Theta, \Omega)$ domains are partitioned using \emph{kd-trees} and scales well to large distributions: partitioning a full SKA dataset usually takes less than 5 minutes.%
\footnote{A full LOFAR dataset is partitioned in a few seconds.}
Once completed, the returned partition map is forwarded to \texttt{allocate()} to configure the transform.
While our proposed auto-chunking approach works well based on our investigations, some $(\Theta,\Omega)$ distributions may achieve better runtimes using hand-tailored partitions.
As such, custom cuts can be provided explicitly to \texttt{allocate()} if desired, side-stepping the auto-chunking approach entirely.
A custom threshold to fallback onto direct evaluation can also be specified to \texttt{allocate()} based on the machine's profile.

\begin{figure}
    \resizebox{\hsize}{!}{\includegraphics{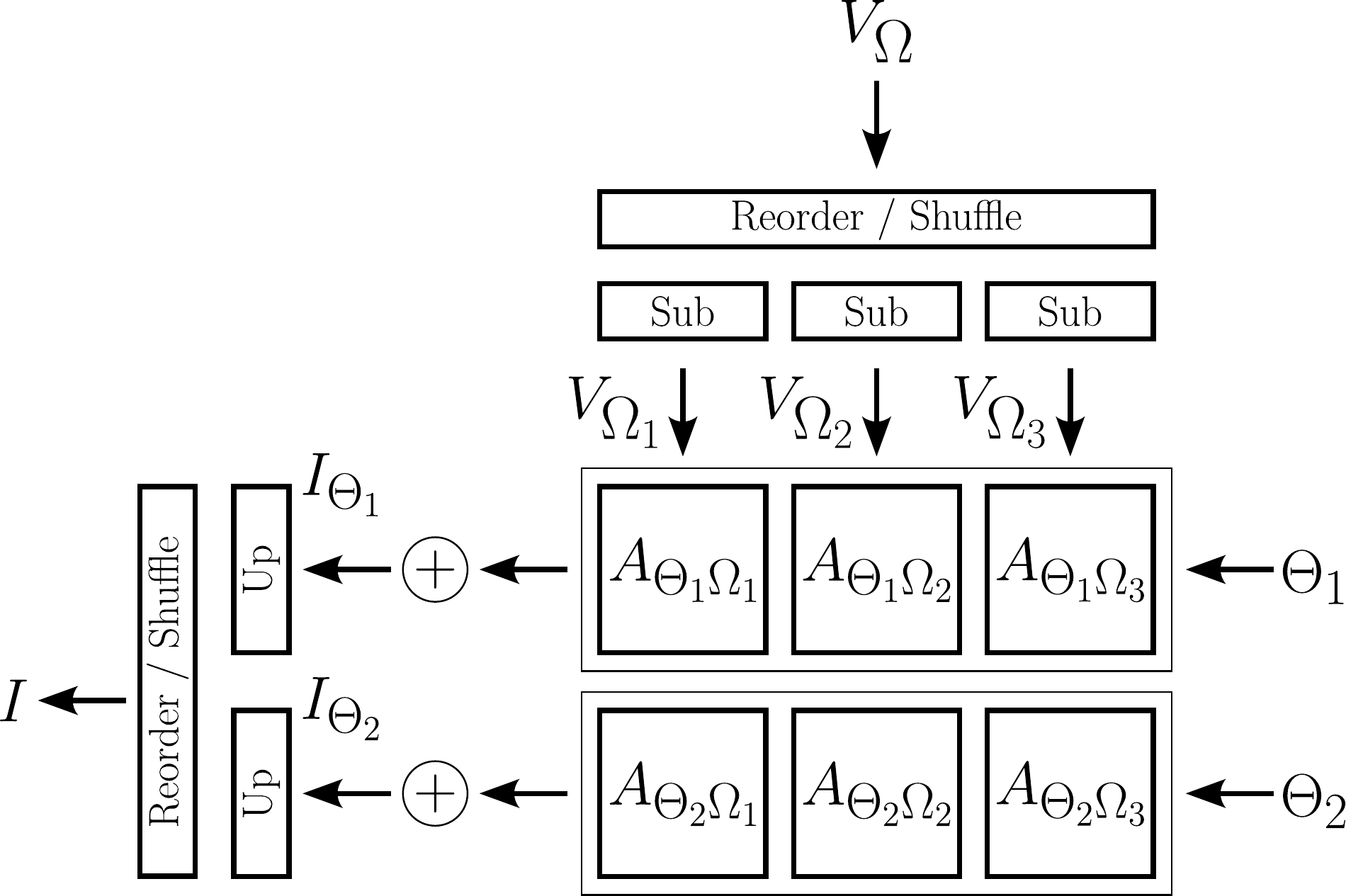}}
    \caption{
        Schematic compute graph to perform \texttt{I = A.adjoint(V)} when $(\Theta,\Omega)$ are partitioned into (2, 3) chunks respectively.
        Inputs $V$ are re-ordered to chunk-order, then sub-sampled to read-only views $\{V_{\Omega_{1}}, V_{\Omega_{2}}, V_{\Omega_{3}}\}$.
        Sub-transforms $(\Theta_{i},\Omega_{j})$ are then computed in sequence, or scheduled for execution via Dask, and summed row-wise prior to writing $I_{\Theta_{i}}$ to the right indices of $I$.
        Computing \texttt{V = A.apply(I)} performs the same steps in reverse order.
    }
    \label{fig:hvox-algo-diag}
\end{figure}

The high-level flow of a chunked transform, i.e.\ a call to \texttt{apply/adjoint()}, is summarized in \figRef{fig:hvox-algo-diag}:
sub-problems collect their respective inputs, perform \tNUFFTthree{} sub-transforms, then sum intermediate products together prior to writing the output.

Collecting inputs amounts to indexing into \texttt{\{x,z,I,V\}} using the built-in \texttt{pycsou.SubSample} operator.
When $(N_{\vis}^{\blk},\,N_{\pix}^{\blk})$ are large however, random indexing multiple times into the above is inefficient.
\texttt{allocate()} therefore re-orders \texttt{\{x,z\}} into chunk-order such that indexing these variables in each sub-problem amounts to slicing contiguous views.
The side-effect is the cost of a permutation of \texttt{\{I,V\}} before/after calls to \texttt{apply/adjoint()} to keep everything coherent.
To avoid paying the same re-ordering cost at each call, it is recommended to supply \texttt{\{x,z,I,V\}} in the "correct" order from the start.
A good re-ordering is computed automatically by \texttt{allocate()} and can be used to initialize a new chunked-transform with better runtime properties if desired.
The overhead of permuting inputs/outputs can therefore be reduced to a pay-once cost when performing a chunked-transform repeatedly.

By default sub-transforms run in sequence to enable imaging large datasets on low-memory systems.
When the system is not memory-starved, it is recommended to set \texttt{parallel=True}, or to feed Dask arrays to \texttt{apply/adjoint()}:
doing so schedules the entire computation as a task graph to execute independent tasks in parallel with significant speedups.

While the memory requirements of individual sub-transforms is capped by \algoRef{algo:box-dim}, the aggregate consumption of all blocks may overwhelm available resources.
\texttt{finufft.Plan}s are therefore created and executed at run-time as needed.
Due to the above, we restrict the FFTW planning effort to \texttt{FFTW\_ESTIMATE}.
Moreover we introduce a process-level lock between sub-transforms when run in parallel because FFTW's planner is not thread-safe.
Since \texttt{finufft.Plan}s cannot be shared among different processes, parallel execution via Dask only works with its thread-based schedulers.%
\footnote{Both the single-machine threaded scheduler and the \texttt{dask.distributed.Client} scheduler in threaded mode are supported.}
Regardless of the above, the FFT's parallelism can be controlled seperately via the \texttt{nthreads} parameter.
We found that setting \texttt{nthreads=<\#physical\_cores>/2} is a good heuristic.

Aside from block-level parallelism, note that the \tFINUFFT{} library used to perform spread/FFT/interpolate steps is itself vectorized:
multiple inputs can thus be transformed simultaneously by controlling the \texttt{ntrans} parameter.
This is particularly useful to image several Stokes parameters in parallel.
Moreover, we note that the \tFINUFFT{} library performs spreading by creating a thread to process \textasciitilde 10k visibilities/pixels \citep{barnett2019parallel}.
Based on our experiments, we found therefore that the unbalanced-chunk problem mentioned in \secRef{subsec:hvox-scale} is not of much concern.

The main drawback of our current implementation is a spread/interpolation overhead due to limitations in \tFINUFFT{}'s Python bindings.
Indeed, its Python API does not allow explicit calls to a type-3's spread, FFT, and interpolation routines.
Moreover, \tFINUFFT{} does not allow users to specify the Heisenberg box dimensions by hand: given a set of \texttt{\{x,z\}} points, \texttt{finufft.Plan} will always use the tightest Heisenberg dimensions possible to achieve the smallest transform.
Put together, these limitations do not allow the overlapping spread/interpolation steps performed by each sub-transform to be factored out at this stage, giving rise to the different complexity of \hvoxI{}(v1) in \tabRef{tb:complexity}.
This will be improved in a future release of \hvoxI{}.

\section{Results}\label{sec:results}
We compare the performance of \hvoxI{} and \twgridder{} in terms of runtime and accuracy.
Both dense and sparse \hpix{}/\dcos{} meshes are investigated at different telescope radii and corresponding image resolutions.
All our experiments build on RASCIL,%
\footnote{
    The Radio Astronomy Simulation, Calibration, and Imaging Library. \url{https://developer.skao.int/projects/rascil/en/latest/index.html}
} maintained and developed by the SDP team.
We use the stable reference implementation of \twgridder{} from the \ducc{} library,%
\footnote{\url{https://gitlab.mpcdf.mpg.de/mtr/ducc}.}
namely the routines \texttt{ducc0.wgridder.[dirty2ms|ms2dirty]()} for analysis and synthesis respectively.%
\footnote{
    The \ducc{} library also contains a more powerful implementation of \twgridder{} under the \texttt{experimental} namespace.
    Being listed as experimental, it was not considered in our benchmarking experiments.
}
Benchmarks were run on a workstation containing an Intel Core i9-10900X 10-core CPU and 128 GB of DDR4 memory.

Our baseline benchmarking setup is described in \tabRef{tb:experiment-defaults}.
When the telescope radius is modified, the number of visibilities and pixel-count associated to each setup is given in \tabRef{tb:experiment1-parameters}.

A gridder's runtime is distributed between spreading, FFT, and (for \hvox{}) interpolation steps.
Large image production is FFT-bound however. (See \secRef{sec:introduction}.)
Given the shared spreading cost between \twgridder{} and \hvox{} (theory, \tabRef{tb:complexity}), the baseline setup is limited to processing a few equi-spaced time-intervals in an 8 hour window to highlight their FFT runtime differences foremost rather than differences in spreader implementations.
That being said, even in this setup, \hvoxI{}v1 incurs a $N_{\pix}^{\blk}$ spreading overhead compared to \twgridder{} due to the limitations mentioned in \secRef{sec:hvox-impl}.
This overhead becomes non-negligeable at large telescope radii where $N_{\pix}^{\blk}$ can be significant, even when choosing a large partition size.
The same holds by analogy for $N_{\vis}^{\blk}$.
Interested readers may refer to \secRef{sec:performance-prediction} where we project via regression analysis the runtime profile of \hvox{} without the spread/interpolation overheads, which matches very closely that of \twgridder{}.
In what follows \hvoxI{} always refers to the v1 implementation, unless said otherwise.

\begin{table}[t!]
	\caption{
        Default parameters used during benchmarks, unless otherwise mentioned.
	}
	\label{tb:experiment-defaults}
	\begin{tabular}{r|l}
		\toprule
		Array                            & SKA-Low                                \\
		Max radius                       & 1 km                                   \\
		Number of antennas               & 251                                    \\
		Number of visibilities           & 94\,878                                \\
        \midrule
		Central frequency                & 150 MHz                                \\
		Bandwidth                        & 50 kHz (1 channel)                     \\
        Phase center                     & RA: $15$\degr, DEC: $-45$\degr         \\
		Field of view                    & $30$\degr                              \\
		Observation duration             & 8 hour (3 time-intervals)              \\
		Image size                       & 0.13 M\pix{} (360 $\times$ 360)        \\
		Angular pixel size               & 5.22\arcsec                            \\
        \midrule
        \hvoxI{} thread-count (chunked)  & 20 (5)                                 \\
		\hvoxI{} max partition size      & 500 MB                                 \\
		\bottomrule
	\end{tabular}
\end{table}

\subsection{Runtime}\label{subsec:results-runtime}
This experiment compares the runtime of \hvoxI{} and \twgridder{} in increasingly demanding imaging configurations given in \tabRef{tb:experiment1-parameters}.
We report the runtime of both analysis (plain lines) and synthesis (dashed lines) codes for (1) \hvox{} when chunking is enabled [\hvoxMB{}] and the monolithic case [\hvox{}]; and (2) \twgridder{} [\nifty{}], including the case where $w$-correction is omitted for faster runs [\niftytd{}].
As a comparison point, we also include the runtime of directly evaluating \eqRef{eq:analysis} and \eqRef{eq:synthesis} via Numba JIT-compiled kernels [DIRECT].

\begin{figure}[t!]
    \resizebox{\hsize}{!}{\includegraphics{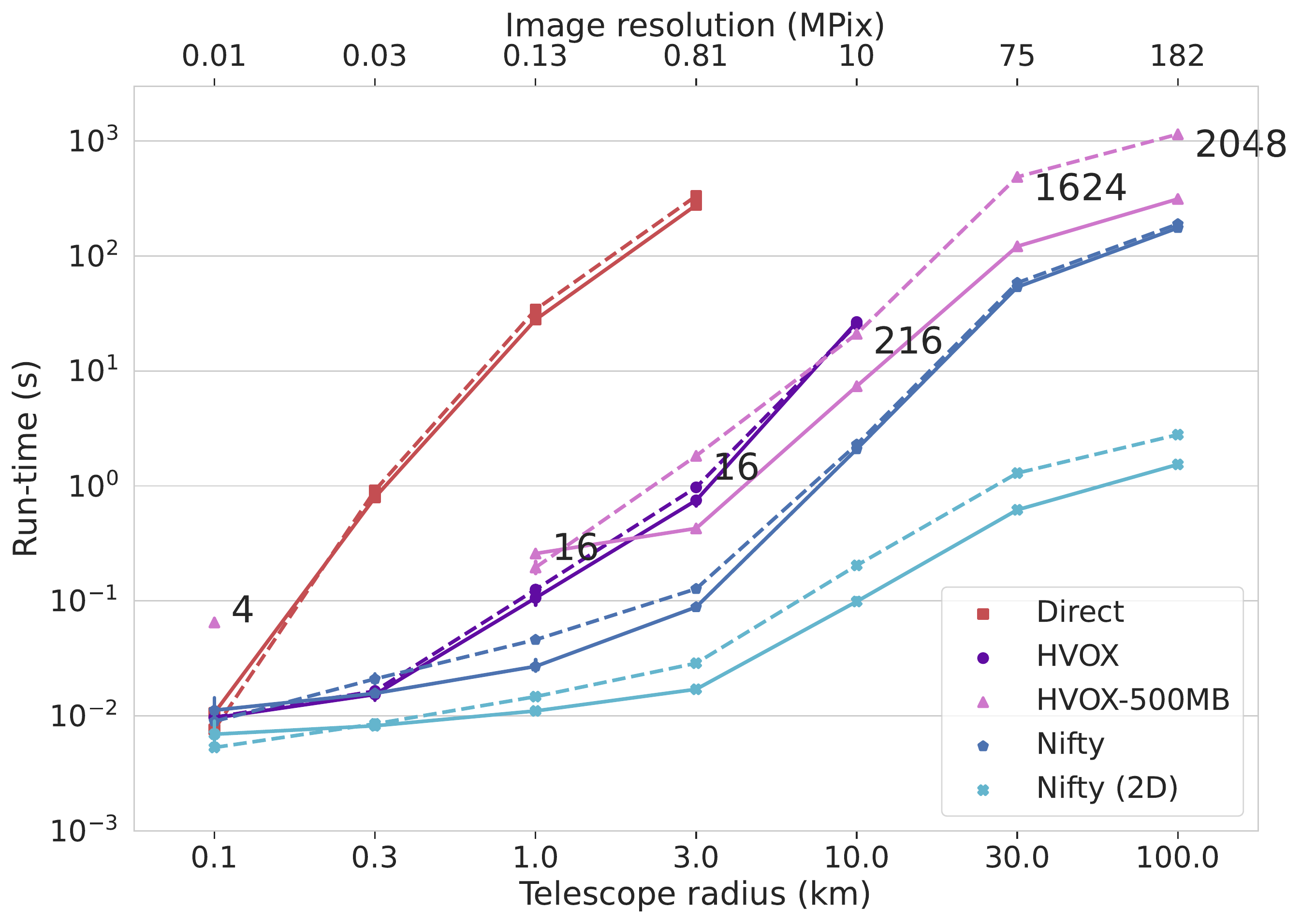}}
    \caption{
        Runtime comparison of analysis (plain lines) and synthesis (dashed lines) for increasing telescope scales. (See \tabRef{tb:experiment1-parameters}.)
        Benchmarked methods are (1) direct evaluation of \eqRef{eq:analysis} and \eqRef{eq:synthesis} [DIRECT], (2) \hvoxI{} without chunking [\hvox{}], (3) \hvoxI{} with at most 500 MB chunks [\hvoxMB{}], (4) \twgridder{} from \ducc{} [\nifty{}, \cite{ye2022high}], and (5) \twgridder{} where no $w$-correction is performed [\niftytd{}].
        The number of chunks used by \hvoxMB{} are indicated as text markers for each configuration.
        Finally, performance of \hvoxMB{} for the 0.3 km configuration is not shown because it resulted in a trivial chunking (only one chunk in each  domain).
    }
    \label{fig:rt-rmax}
\end{figure}
\begin{figure}[t!]
    \resizebox{\hsize}{!}{\includegraphics{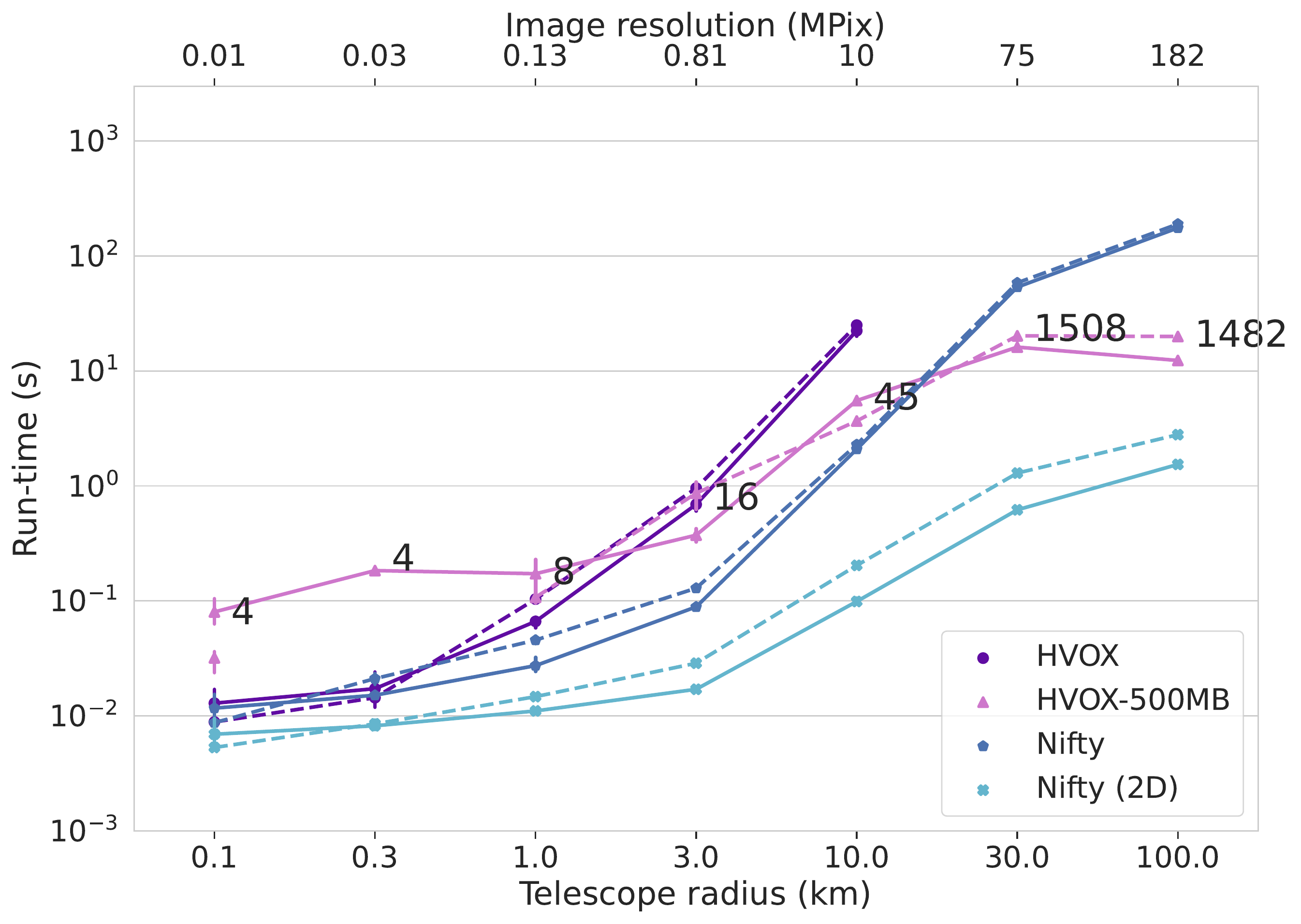}}
    \caption{
        Runtime comparison of analysis (plain lines) and synthesis (dashed lines) for increasing telescope scales on sky maps where only 500 pixels are of interest.
        The \nifty{} and \niftytd{} lines are the same as in \figRef{fig:rt-rmax} since these methods cannot cater to irregular meshes, and thus image sparse domains.
    }
    \label{fig:rt-rmax-sparse}
\end{figure}

\figRef{fig:rt-rmax} shows the runtimes of aforementioned methods to perform analysis and synthesis from/to a \dcos{} mesh of size $N_{\pix}$ suitable for post-processing applications such as deconvolution. (\tabRef{tb:experiment1-parameters}.)
The results are on par with the arithmetic complexities provided in \tabRef{tb:complexity}.
With its bi-linear complexity, direct evaluation is systematically orders of magnitude slower than FFT-based methods, except for small array configurations.
Amongst FFT-based methods, \niftytd{} is obviously the fastest, in particular for large imaging problems where it outperforms other methods at the cost of low accuracy. (See \secRef{subsec:results-accuracy}.)
When performing $w$-correction [\nifty{}], \twgridder{} performs better than \hvox{} methods, especially in the absence of chunking and for mid-scale setups (\textasciitilde 1--10 km radii).
For larger configurations, \hvox{} (monolithic) runs out of memory due to the size of the Heisenberg box required to bound all visibilities, as predicted and discussed in \secRef{subsec:hvox-scale}.
In contrast, \hvoxMB{} scales nicely to such setups and even reduces the performance gap with respect to \twgridder{}: both methods exhibit similar runtimes for analysis, despite \hvoxMB{} having an extra interpolation step.
This is particularly encouraging given that \hvoxI{}v1 is mainly Python-based, contains no domain-specific optimizations, and has a spread/interpolation overhead compared to \twgridder{}. (See \tabRef{tb:complexity}.)

In synthesis mode, all methods incur a performance drop due to the parallelism asymmetry between spreading and interpolation steps in NUFFT algorithms \citep{barnett2019parallel}.
The performance drop is very noticeable for \hvoxMB{} due to the spreading/interpolation overhead of \hvoxI{}v1.

Not all applications require performing analysis or synthesis on a dense \dcos{} grid:
direction-dependant calibration for example consists in evaluating these operators in the vicinity of known calibrators to estimate, then correct for, atmospheric-induced distortions.
\figRef{fig:rt-rmax-sparse} shows the runtimes of analysis and synthesis using \hvox{} and \twgridder{}-based methods from/to a \emph{sparse} \dcos{} mesh matching this context, namely when only small (not necessarily contiguous) regions of the grid are to be analyzed/synthesized respectively.
The results show that \hvoxMB{} performs both steps one order of magnitude faster than \nifty{} despite aforementioned overheads.
This is because chunked evaluation performs analysis/synthesis only on the regions of interest, whereas \twgridder{} must process the whole image independently of its underlying sparsity.
This illustrates the strong potential of \hvoxI{} when sparse sky/visibility domains are involved, as often encountered in CLEAN iterations or during calibration loops.

\begin{figure}[t!]
    \resizebox{\hsize}{!}{\includegraphics{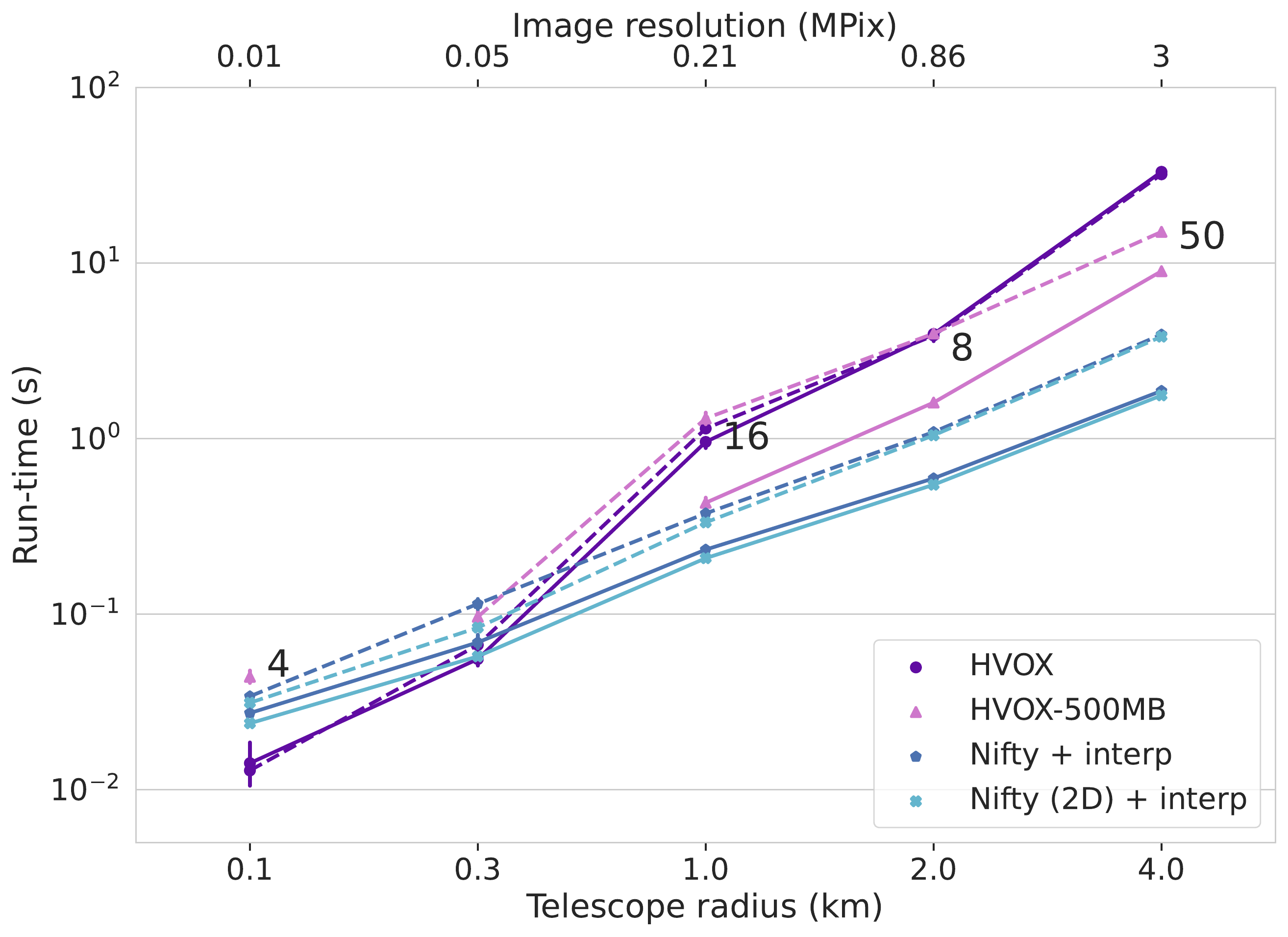}}
    \caption{
        Runtime comparison of analysis (plain lines) and synthesis (dashed lines) for increasing telescope scales on \hpix{} skies.
        Analysis/synthesis using \twgridder{}-based methods [\nifty{}, \niftytd{}] are performed on the \dcos{} mesh, but preceded/followed by interpolation from/to a suitably-scaled \hpix{} mesh.
        Runtimes of \hvox{}-based methods [\hvox{}, \hvoxMB{}] already include the built-in interpolation time.
    }
    \label{fig:rt-rmax-hpix}
\end{figure}

\begin{figure}[t!]
    \resizebox{\hsize}{!}{\includegraphics{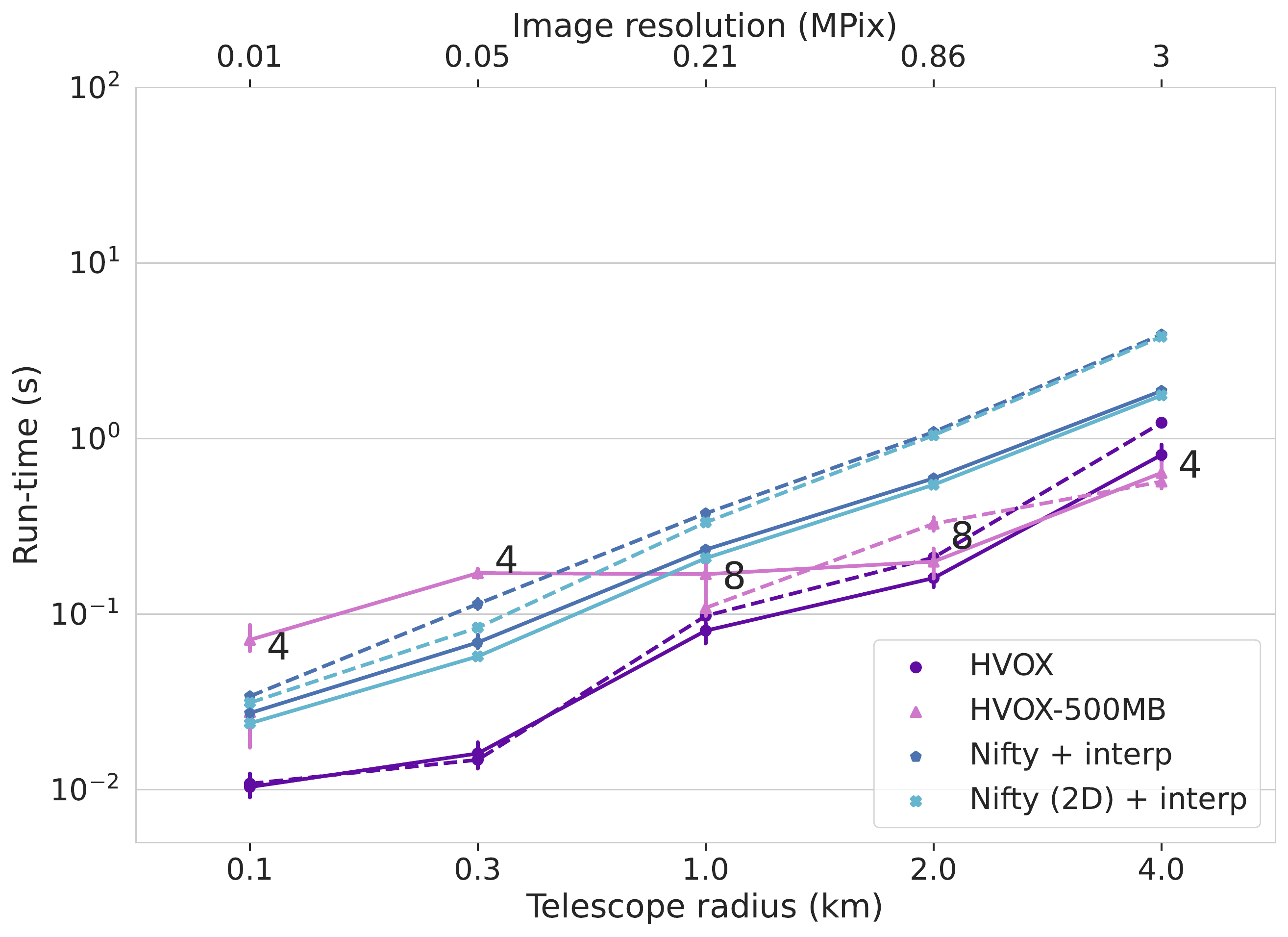}}
    \caption{
        Runtime comparison of analysis (plain lines) and synthesis (dashed lines) for increasing telescope scales on sparse \hpix{} skies where only 500 pixels are of interest.
        Analysis/synthesis using \twgridder{}-based methods [\nifty{}, \niftytd{}] are performed on the \dcos{} mesh, but preceded/followed by interpolation from/to a suitably-scaled \hpix{} mesh.
        Runtimes of \hvox{}-based methods [\hvox{}, \hvoxMB{}] already include the built-in interpolation time.
    }
    \label{fig:rt-rmax-hpix-sparse}
\end{figure}

One of \hvox{}'s main benefits is the ability to perform mesh-agnostic analysis and synthesis.
This property is not shared by \twgridder{}-based methods:
if a specific mesh is required for post-processing tasks such as harmonic analysis, then the \dcos{} mesh returned/expected by \twgridder{} must be re-interpolated from/to the suitable mesh beforehand.
This section concludes by assessing the overhead incurred by interpolating \dcos{} meshes to \hpix{} specifically.
To do so, the runtime of \hvox{}-based methods are compared against \twgridder{}-based ones described above, where \twgridder{} runs are preceded/followed by a suitable re-interpolation from/to the supported \dcos{} mesh.
This additional step is performed using optimized bi-linear interpolation routines.%
\footnote{
    We use AstroPy-HEALPix's \texttt{interpolate\_bilinear\_skycoord()} method to interpolate from \hpix{} to \dcos{} \citep{astropy:2013,astropy:2018,astropy:2022},
    and \texttt{scipy.interpolate.interpn()} for the converse \citep{virtanen2020scipy}.
}
\figRef{fig:rt-rmax-hpix} and \figRef{fig:rt-rmax-hpix-sparse} show the obtained runtimes for dense and sparse \hpix{} meshes respectively, by analogy with \figRef{fig:rt-rmax} and \figRef{fig:rt-rmax-sparse}.
As outlined in \secRef{sec:introduction}, the large cost of spherical re-interpolation completely masks \niftytd{}'s runtime advantage with respect to \nifty{}.
Moreover, while \twgridder{} methods [\nifty{}, \niftytd{}] still outperform \hvox{} methods [\hvox{}, \hvoxMB{}] in the dense case, \hvox{} significantly outperforms \twgridder{} methods in the context of sparse meshes.
Performing image formation directly in the mesh of interest is thus computationally attractive.
\begin{table*}[t!]
	\centering
	\caption{
		Imaging configurations used for \secRef{subsec:results-runtime} experiments.
        The associated telescope layouts are shown in \figRef{fig:skabd2-antenna-dist}.
	}
	\label{tb:experiment1-parameters}
	\begin{tabular}{l|ccccccc}
		\toprule
		Telescope radius [km] & 0.1    & 0.3    & 1    & 3    & 10    & 30    & 100    \\
        \midrule
		$N_{\vis}$            & 273    & 13k    & 95k  & 148k & 238k  & 341k  & 394k   \\
		$N_{\pix}\; [M\pix]$  & 0.0064 & 0.0256 & 0.13 & 0.81 & 10.24 & 74.65 & 182.25 \\
        \bottomrule
	\end{tabular}
\end{table*}

\subsection{Accuracy}\label{subsec:results-accuracy}
As outlined in \secRef{sec:introduction}, image formation is often followed by image-processing pipelines to extract useful information from data cubes.
However these tasks perform poorly when applied to \dcos{} meshes, reason why interpolating to a suitable mesh is necessary prior to any processing.
Spherical interpolation is expensive however (\figRef{fig:rt-rmax-hpix} and \figRef{fig:rt-rmax-hpix-sparse}) and breaks any accuracy guarantees provided by gridders.
To quantify the accuracy loss, this experiment compares the normalized mean square error (\nmse{}) between ground-truth visibilities $V_{\Omega}^{GT}$ obtained via direct evaluation of a point-source sky model defined on a mesh $\Theta$, to those produced by \hvox{}-based methods [\hvox{}, \hvoxMB{}] and \twgridder{}-based methods [\nifty{}, \niftytd{}].

\begin{figure}[t!]
    \begin{center}
        \resizebox{0.45\hsize}{!}{\includegraphics{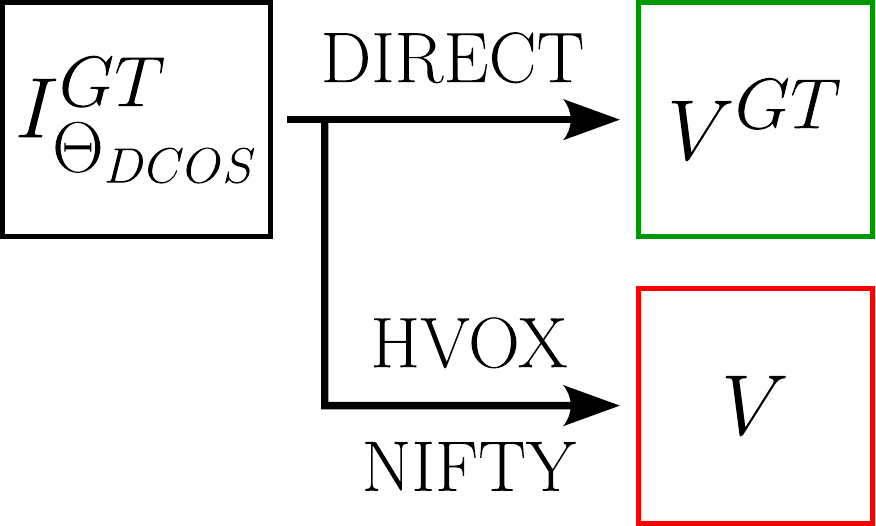}}
    \end{center}
    \resizebox{\hsize}{!}{\includegraphics{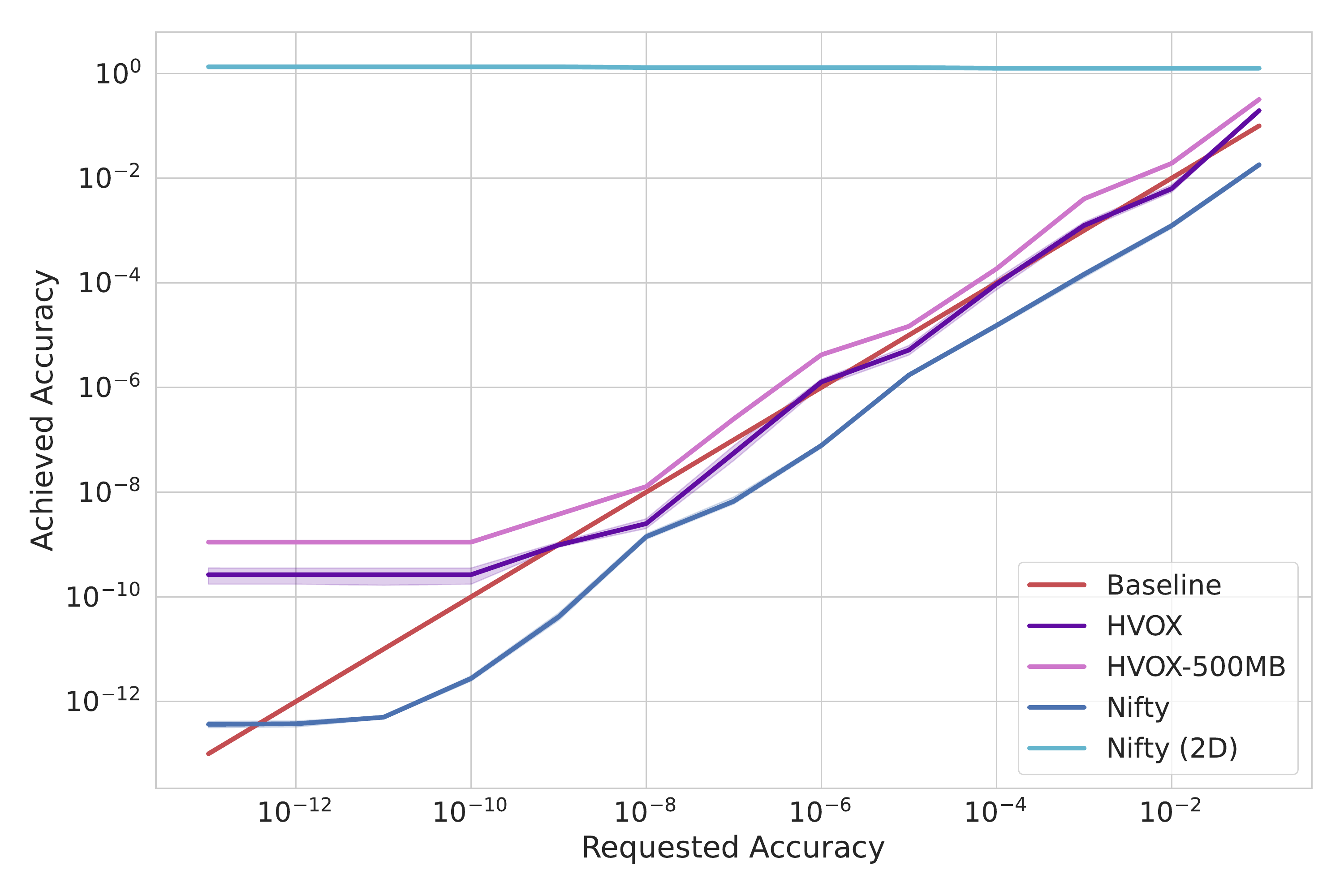}}
    \caption{
        Accuracy compliance of gridder implementations for \dcos{} meshes. \\
        \textbf{Top:} ground-truth visibilities $V^{GT}$ produced by direct-evaluation of a point source sky model $I_{\Theta_{\dcos{}}}^{GT}$ defined on a \dcos{} grid [green] are compared with visibilities $V$ [red] produced by \hvox{}-based [\hvox{}, \hvoxMB{}] and \twgridder{}-based implementations [\nifty{}, \niftytd{}]. \\
        \textbf{Bottom:} comparison of achieved \nmse{} accuracy (vertical axis) given a requested accuracy target (horizontal axis).
        A baseline is included to represent the identity relationship.
    }
    \label{fig:acc-eps}
\end{figure}

When the sky model is defined in a \dcos{} mesh $\Theta_{\dcos{}}$, then both \nifty{}- and \hvox{}-based methods perform analysis directly from $\Theta_{\dcos{}}$ to visibility space $\Omega$ as depicted in \figRef{fig:acc-eps}[top], with associated \nmse{} mismatch in \figRef{fig:acc-eps}[bottom].
We observe that the requested accuracy is closely provided by \hvox{} (monolithic) up to a precision of $10^{-9}$ only.
This can be explained by the fact that FINUFFT uses by default a low upsampling factor of 1.25 for type-3 transforms, allowing for lower RAM usage and smaller FFTs but limiting the achievable accuracy to 9 digits.%
\footnote{See FINUFFT's online documentation regarding the \texttt{upsampfac} optional parameter: \url{https://finufft.readthedocs.io/en/latest/opts.html\#algorithm-performance-options}.}
The \hvoxMB{} curve closely follows \hvox{}, but slightly under-achieves the accuracy target due to accumulation of numerical error by adding chunks together.
This effect can be compensated for by requesting a smaller accuracy target $\epsilon$ (e.g., dividing it by the total number of blocks) for the sub-transforms involved in chunked evaluation.
Turning to \twgridder{}-based methods, \nifty{} systematically overachieves the requested accuracy by one order of magnitude, a behaviour consistent with the results of \citep{arras2021efficient}.
This suggests that \nifty{} may be doing slightly more computational work than necessary to achieve a requested accuracy.
On the other hand \niftytd{}, despite its runtime efficiency (\secRef{subsec:results-runtime}), exhibits a trade-off between accuracy and runtime by not accounting for the W-term.
As a consequence, it suffers from significantly diminished accuracy compared to \nifty{}, independent of the requested target.

\begin{figure}[t!]
    \begin{center}
        \resizebox{0.45\hsize}{!}{\includegraphics{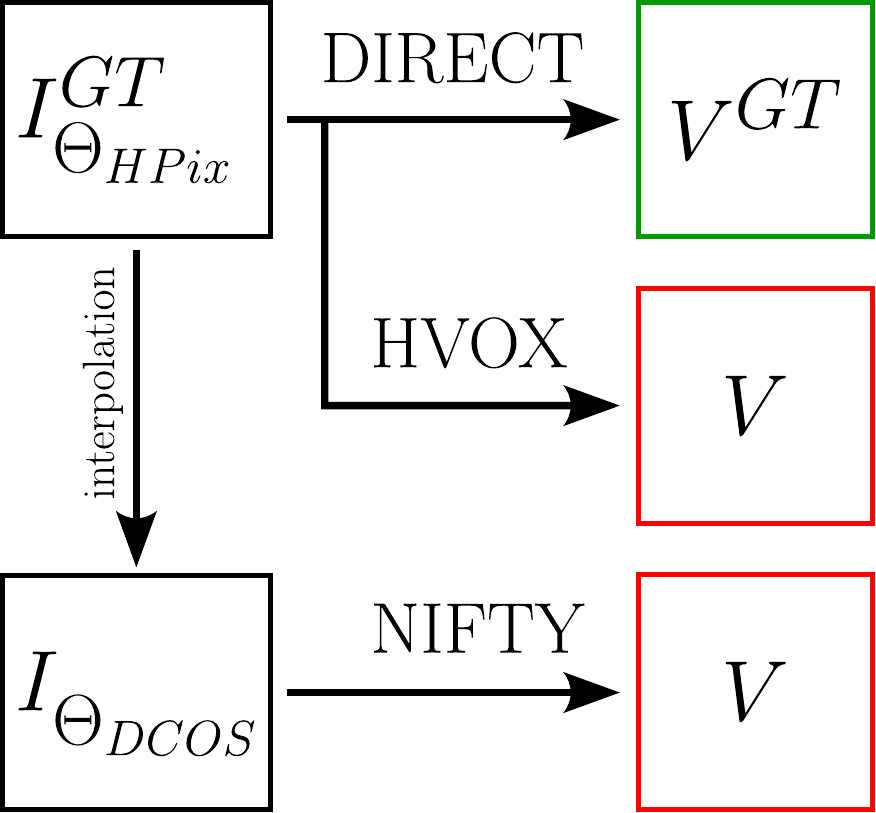}}
    \end{center}
    \resizebox{\hsize}{!}{\includegraphics{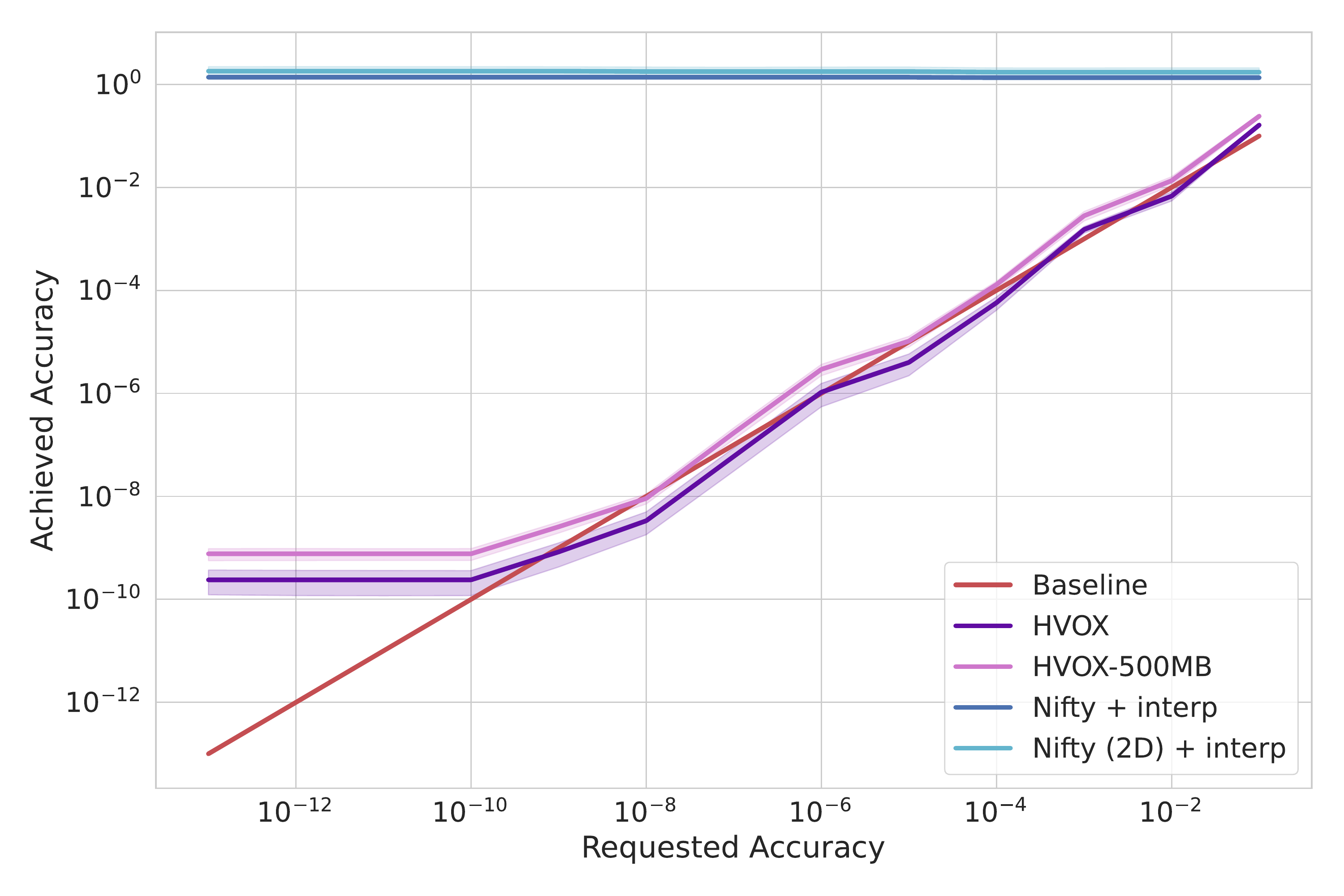}}
	\caption{
        Accuracy compliance of gridder implementations for \hpix{} meshes. \\
        \textbf{Top:} ground-truth visibilities $V^{GT}$ produced by direct-evaluation of a point source sky model $I_{\Theta_{\hpix{}}}^{GT}$ defined on a \hpix{} grid [green] are compared with visibilities $V$ [red] produced by \hvox{}-based [\hvox{}, \hvoxMB{}] and \twgridder{}-based implementations [\nifty{}, \niftytd{}] after interpolation. \\
        \textbf{Bottom:} comparison of achieved \nmse{} accuracy (vertical axis) given a requested accuracy target (horizontal axis).
        A baseline is included to represent the identity relationship.
    }
	\label{fig:acc-eps-hpix}
\end{figure}

When the sky model is defined in a \hpix{} mesh $\Theta_{\hpix{}}$ however, then \twgridder{}-based methods cannot perform analysis directly:
an interpolation is required beforehand to map $\Theta_{\hpix{}}$ to an equivalent \dcos{} mesh $\Theta_{\dcos{}}$ supported by \nifty{} and \niftytd{}.
The processing chain is outlined in \figRef{fig:acc-eps-hpix}[top], with associated \nmse{} mismatch in \figRef{fig:acc-eps-hpix}[bottom].
The results show that \hvox{}-based methods perform identically irrespective of the source mesh $\Theta$.
\twgridder{}-based methods on the other hand suffer an important loss of accuracy due to interpolating $\Theta_{\hpix{}}$ to $\Theta_{\dcos{}}$ prior to performing analysis.
The magnitude of this interpolation error is significant ($\sim 1 \%$), to the point that the achieved accuracy is similar to not accounting for the W-term at all (and that independently from the requested accuracy).
Direct inclusion of the effect of interpolation during gridding, as \hvox{} does, thus allows for analysis and synthesis steps with \emph{controllable} accuracy irrespective of mesh constraints.
This may reveal extremely important in certain scientific test cases.

\begin{figure}
	\resizebox{\hsize}{!}{\includegraphics{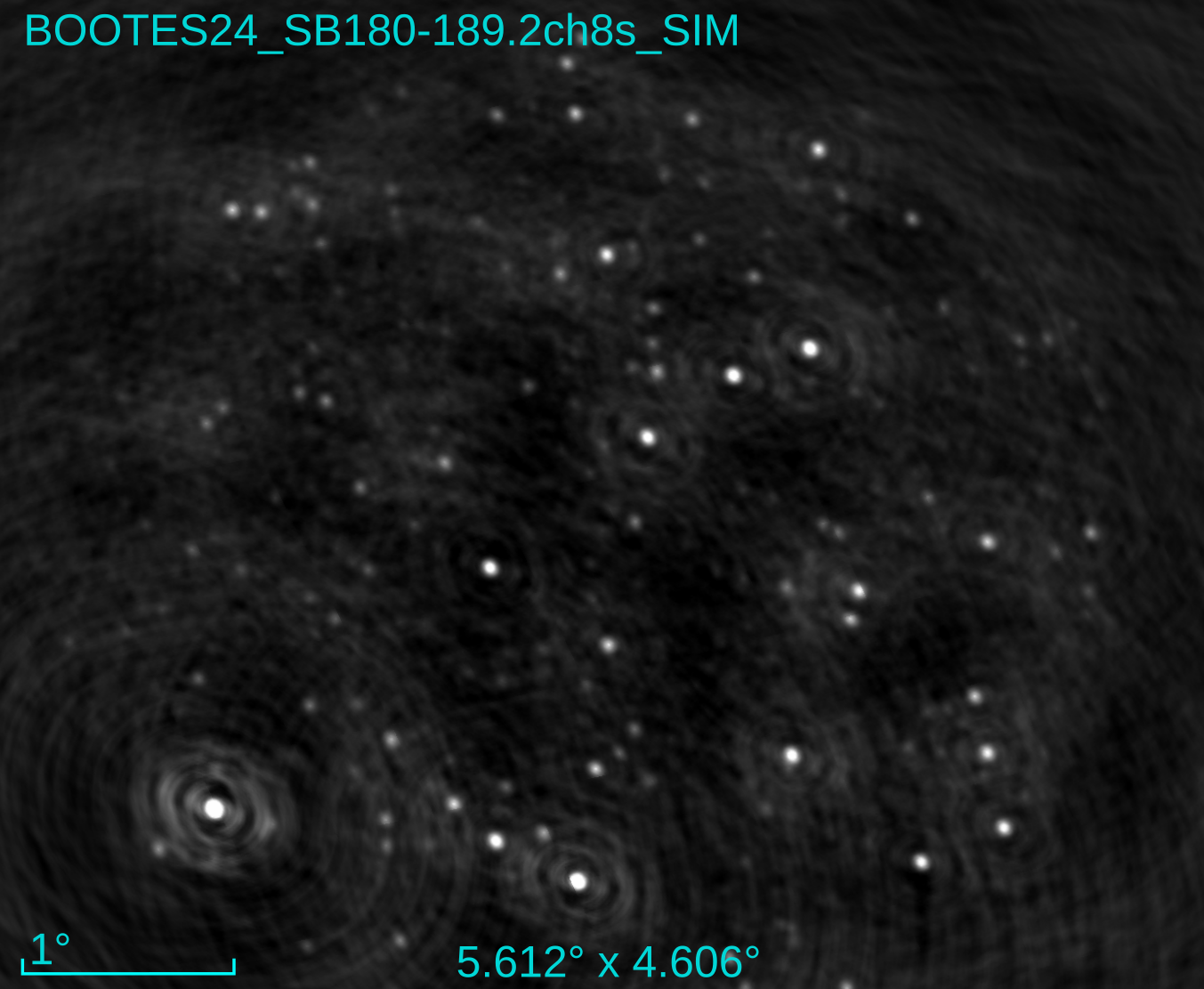}}
	\caption{
        Dirty \hpix{} image of an 8 hour observation of the Boötes field taken with LOFAR-HBA (24 core stations \& 14 remote stations).
        The image was produced by loading a LOFAR measurement set using RASCIL, imaging the $5\degr \times 5\degr$ field on a \textasciitilde 1.5 M\pix{} \hpix{} mesh using \hvoxMB{}, then writing the image to a FITS file.
        The image was generated on the workstation described in \secRef{sec:results} in 38 seconds.
        Performing synthesis onto a \hpix{} mesh thus reduces to changing a few lines in a RASCIL script.
    }
	\label{fig:bootes-hpix}
\end{figure}

\section{Conclusion and future perspectives}\label{sec:future}
Fast, accurate, and scalable analysis and synthesis are key to achieving the image processing goals of the SKA.
This paper introduced the \emph{Heisenberg voxelisation} method (\hvox{}), a novel approach to perform these compute-critical steps.
By performing analysis and synthesis with a type-3 NUFFT in 3D, \hvox{} allows image production onto an arbitrary pixel mesh.
Images can thus be produced directly in a suitable mesh for image post-processing pipelines with strong accuracy guarantees.
Moreover \hvox{} is inherently scalable by decoupling the 3D FFT size from the image size.
Our chunking approach further increases scalability by leveraging the local structure of baselines/mapped sky regions to drastically reduce the FFT memory requirements, execute chunks in parallel, and tailor the synthesis method at the chunk-level for increased performance.
We also provide \hvoxI{}, the first CPU-based implementation of the \hvox{} method based on the PyData stack and the FINUFFT type-3 engine.
With its high-level API, \hvoxI{} can be used as a drop-in replacement for current \dcos{} gridders in RASCIL, easing adoption and deployment.
We refer interested users to the \hvoxI{} repository\footnote{\url{https://github.com/matthieumeo/hvox}} containing example codes to produce \hpix{} images from MS files, allowing production of images such as \figRef{fig:bootes-hpix}.

Despite the mostly Python-based nature of our implementation, the lack of domain-specific optimizations, and the algorithmic overhead incurred by \hvoxI{}v1, our results show that \hvox{} is runtime competitive with \twgridder{}-based codes for high-resolution wide-field imaging.
In the case of sparse fields/baselines moreover, \hvox{} is able to fully leverage this structure via its chunked-based evaluation strategy to outperform \twgridder{}.

Accuracy-wise, \hvox{} consistently achieves specified accuracy targets up to $10^{-9}$, irrespective of the mesh used.
In contrast \twgridder{}'s accuracy guarantees are lost once interpolation to a \hpix{} mesh is required during post-processing pipelines.

While our implementation has a lot of room for improvement, we hope to have conveyed the importance of \tNUFFTthree{}-based analysis and synthesis to the community.
Future work includes (1) factoring out the spread/interpolation steps from sub-transforms to drop current overheads in \hvoxI{}v1, (2) adding support for beamshapes and direction-dependant effects, and (3) providing a GPU implementation of \hvox{}.
Beyond \hvox{}-proper, we look forward to the novel applications enabled by the method, notably improved calibration via its ability to perform fast synthesis of sparse sky maps.
In addition we forsee great potential in reconsidering image reconstruction methods which are sparsity-aware \citep{jarret2022fast}, and stochastic optimization in general.
Finally, having decoupled FFT memory requirements from image size, we ponder on the potential implications of \hvox{} on large-scale image processing systems such as the SDP.

\begin{acknowledgements}
    Sepand Kashani and Adrian Jarret were funded by the Swiss National Science Foundation (SNSF) under grant 200021 181978/1 "SESAM -- Sensing and Sampling: Theory and Algorithms".
    Adrian Jarret was also funded by the SNSF grant CRSII5\_193826 "AstroSignals".
    Matthieu Simeoni and Joan Rué Queralt were funded by SNSF grant CRSII5\_193826 "AstroSignals" and the State Secretariat for Education Research and Innovation (SERI) as part of the SKA Switzerland Consortium (SKACH).
    The authors are grateful to Martin Vetterli, Paul Hurley, Julien Fageot, and Martin Reinecke for their useful feedback and numerous advice on earlier versions of this work.
    Their expert advice has largely contributed to improving the overall quality of this manuscript.
\end{acknowledgements}

\bibpunct{(}{)}{;}{a}{}{,} 
\bibliographystyle{aa}
\bibliography{references}

\begin{appendix}
    \section{\hvoxI{} (theory) performance prediction}\label{sec:performance-prediction}
    Due to constraints imposed by the \tFINUFFT{} library used in \hvoxI{} to perform \tNUFFTthree{} transforms, \hvoxI{}v1 has a spread/interpolation overhead compared to \hvox{} (theory, see \tabRef{tb:complexity}.)
    The benchmark setup (\tabRef{tb:experiment-defaults}) chosen for runtime results in \secRef{sec:results} highlights FFT runtime differences foremost, however the spread/interpolation overhead is still non-negligeable when the telescope radius is increased.
    The overhead is even more pronounced if all time intervals are used.
    The true strength of the \hvox{} method is thus not shown in \secRef{sec:results}, especially when performing analysis from (or synthesis to) \emph{dense} meshes, i.e.\ \figRef{fig:rt-rmax} and \figRef{fig:rt-rmax-hpix}.

\begin{figure}[t!]
        \resizebox{\hsize}{!}{\includegraphics{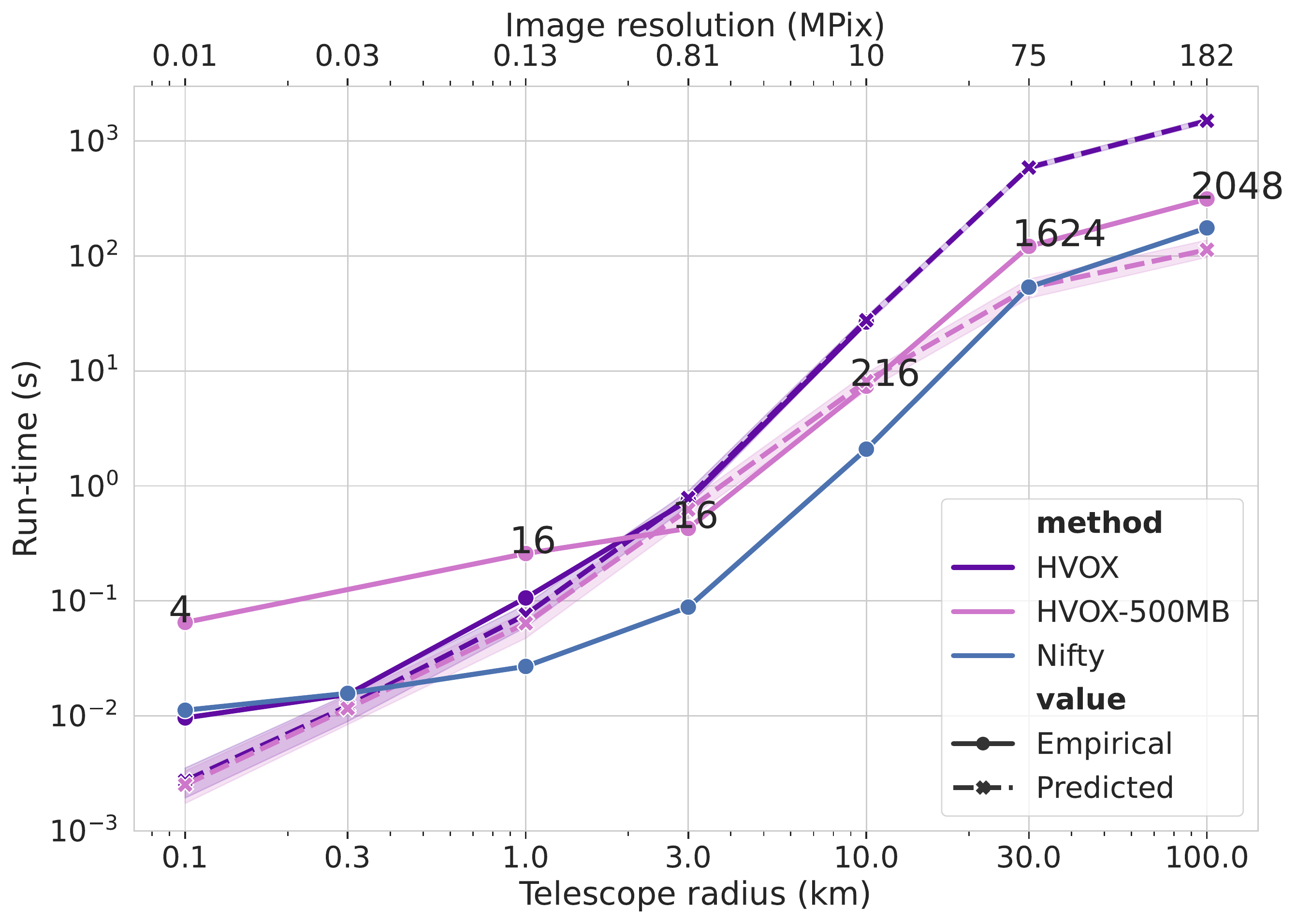}}
        \caption{
            Predicted runtime to perform analysis from a dense \dcos{} mesh using \hvox{} and \hvoxMB{} when the spread/interpolation of \hvoxI{}v1 is removed.
            Predicted runtimes are obtained via regression analysis. (See \secRef{sec:performance-prediction}.)
        }
        \label{fig:prediction}
    \end{figure}

    To get a better glimpse at the achievable performance of \hvox{}, we built a regression model of \hvox{} (monolithic) runtimes using different ($N_{\vis}$, $N_{\pix}$, $\epsilon$) pairs.
    This resulted in a total of 320 \hvox{} runtimes which we used to perform a linear regression and determine the complexity contribution factors of the spread, FFT, and interpolation steps, as shown in \tabRef{tb:complexity}.
    Using the regressed parameters and their (5, 95) confidence intervals, we predicted the runtime of \hvox{} with the partitioning scheme in an ideal scenario where redundant spreading and interpolation are eliminated.
    The results, depicted in \figRef{fig:prediction}, indicate that \hvox{} has the potential to achieve faster runtimes than \nifty{} for dense meshes, showcasing its capability for further improvement.


    \begin{figure}
    \resizebox{\hsize}{!}{\includegraphics{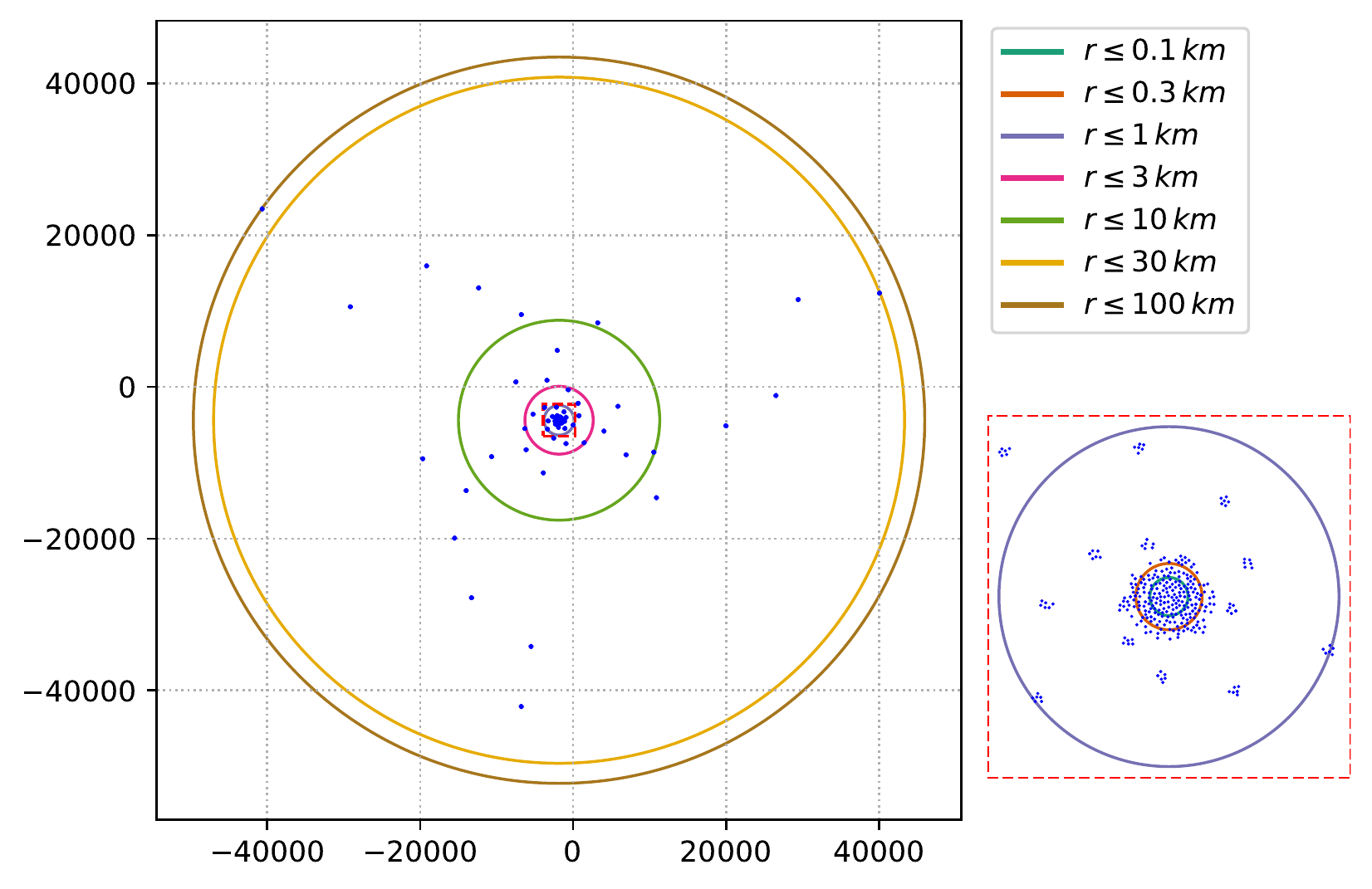}}
    \caption{
        SKA-LOW antenna geometry used in \secRef{subsec:results-runtime}.
        The 512-element array consists of a dense core, with remaining antennas spiraling out to vastly increase the array aperture.
        The inset shows a zoom-in on antennas within 1 km of the array center.
    }
    \label{fig:skabd2-antenna-dist}
\end{figure}

\end{appendix}

\end{document}